\documentclass[twocolumn]{aastex62}
\usepackage[utf8]{inputenc}
\usepackage[T1]{fontenc}
\usepackage{amsmath}

\newcommand{\sao}{\affiliation{Center for Astrophysics $\vert$ Harvard \& Smithsonian, Cambridge, MA, USA}}
\newcommand{\harvard}{\affiliation{Department of Physics, Harvard University, Cambridge, MA, USA}}
\newcommand{\apll}{\affiliation{Johns Hopkins Applied Physics Laboratory, Laurel, MD 20723, USA}}
\newcommand{\gsfc}{\affiliation{Heliophysics Science Division, NASA, Goddard Space Flight Center, Greenbelt, MD 20771, USA}}
\newcommand{\ssl}{\affiliation{Space Sciences Laboratory, University of California, Berkeley, CA, USA}}
\newcommand{\iowa}{\affiliation{Department of Physics and Astronomy, University of Iowa, Iowa City, IA, 52242, USA}}

\shorttitle{Radial Evolution of Solar Wind streams}
\shortauthors{Ahmed et al. 2026}

\begin{document}

\title{Tracing the Radial Evolution of Switchback Patches into Microstreams in Alfv\'enic Slow Solar Wind Using Physics Informed Dynamic Time Warping (Pi-DTW)} 
\correspondingauthor{Lidiya Ahmed}
\email{lidiya\_ahmed@g.harvard.edu}

\author[0000-0002-2559-0831]{Lidiya Y. Ahmed}\harvard 
\author[0000-0002-7728-0085]{Michael L Stevens}\sao
\author[0000-0002-6145-436X]{Samuel T. Badman}\sao
\author[0000-0002-8748-2123]{Yeimy J. Rivera}\sao
\author[0000-0003-0896-7972]{Srijan Bharati Das}\sao
\author[0000-0002-5699-090X]{Kristoff Paulson}\sao
\author[0000-0003-2409-3742]{Nour E. Raouafi}\apll
\author[0000-0001-6673-3432]{B.\ L.\ Alterman}\gsfc
\author[0000-0002-1989-3596]{Stuart D. Bale}\ssl
\author[0000-0001-5030-6030]{Davin E. Larson}\ssl
\author[0000-0003-0519-6498]{Ali Rahmati}\ssl
\author[0000-0002-0396-0547]{Roberto Livi}\ssl
\author[0000-0002-7287-5098]{Phyllis Whittlesey}\ssl
\author[0000-0001-5258-6128]{Jasper S. Halekas}\iowa

\begin{abstract}
Switchback (SB) patches observed near the Sun and microstreams observed farther out in the heliosphere are ubiquitous mesoscale solar wind structures with similar angular extents of a few degrees. 
This similarity raises the hypothesis that SB patches evolve into microstreams as they propagate away from the Sun. Testing this hypothesis is important for understanding the evolution of these energy containing structures in the young solar wind and may enable finer-scale prediction of solar wind arriving at Earth. We use a unique observational alignment between Parker Solar Probe ({\it{Parker}}), Solar Orbiter (SolO), and \textit{Wind} during {\it{Parker}}'s Encounter 19, covering solar distances from 11 to 210 $R_{\odot}$, in the lead up to the April 8, 2024 total solar eclipse over North America. We use Physics Informed Dynamic Time Warping (Pi-DTW), incorporating conserved quantities into the cost function and using heliospheric current sheet crossings as alignment markers, to associate solar wind structures across heliocentric distances and investigate their radial evolution. We find that: (1) Using our Pi-DTW method, the same mesoscale plasma structures can be traced between spacecraft despite undergoing stretching and compression; and (2) the observed structures are generally consistent with previously reported microstream characteristics and angular scales ($5$-$10$\textdegree), with generally decreasing Alfv\'enicity and magnetic fluctuation power while mesoscale velocity enhancements remain identifiable through SolO and \textit{Wind}. Overall, our observations are consistent with the hypothesis that SB patches evolve into microstream like mesoscale velocity structures. 
\end{abstract}


\section{Introduction}
\label{sec:intro}

The solar wind, a continuous stream of hot plasma, carries the interplanetary magnetic field from the Sun throughout the heliosphere. Outflows from the Sun drive space weather that can affect Earth’s magnetosphere, and therefore power grids, satellites, GPS, Auroras, and other planetary magnetospheres. The solar wind has been studied for decades, and despite the significant contributions made by recent studies using Parker Solar Probe \citep[hereafter {\it{Parker}};][]{Fox2016,Raouafi2023}, and Solar Orbiter (SolO) \citep{20_Muller2020} observations, several fundamental questions about how solar wind structures evolve as they propagate away from the Sun remain unresolved. 

\begin{figure*}[ht!]
\centering
\includegraphics[width=1.0\linewidth]{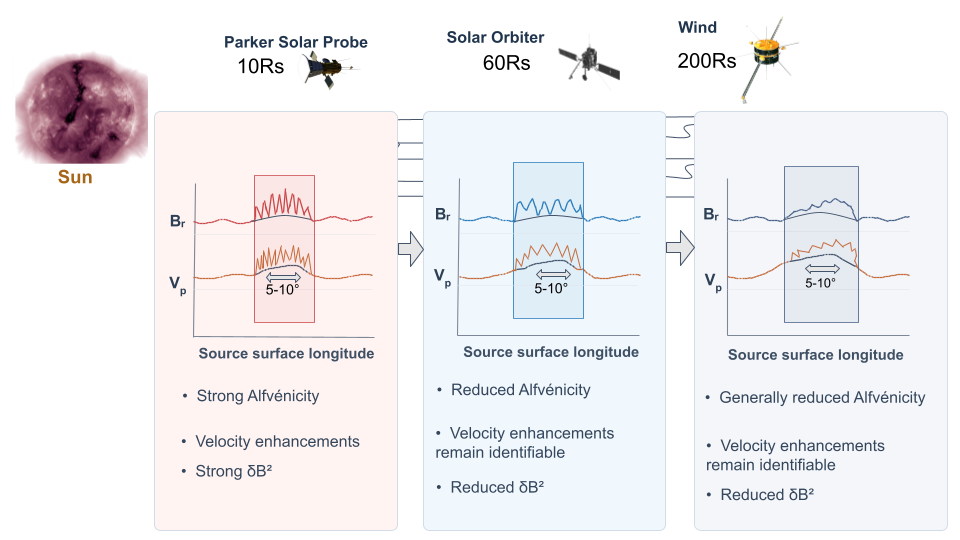}
\caption{A Schematic illustration of the proposed radial evolution of switchback (SB) patches into microstream like structures. Near Parker Solar Probe, SB patches exhibit strong Alfv\'enicity, large magnetic field fluctuations, and velocity enhancements. With increasing heliocentric distance, the magnetic fluctuation power ($\delta B^2$) and Alfv\'enicity generally decrease, while mesoscale velocity enhancements remain identifiable at Solar Orbiter and \textit{Wind}. The schematic illustrates the evolutionary scenario investigated in this study.}
\label{Figure 1.}
\end{figure*}

Magnetic switchbacks (SBs), Alfvénic fluctuations observed as synchronous fluctuations in the solar wind's magnetic field and velocity that maintain constant magnetic field strength \citep{Kasper_2019Nature, Bale_2019Nature, Bale_2021}, have been observed in abundance near the Sun by {\emph{Parker}}. Recent observations have shown that SBs are organized into larger SB patches spanning angular scales of a few degrees, typically of order 5-10\textdegree, comparable to scales associated with solar supergranulation \citep{Horbury2023}. SB patches are characterized as ``arc-like'' structures in the solar wind's  background radial magnetic field strength and velocity, with individual SBs presenting as 1 sided large deflections from these arcs (Fig. \ref{Figure 1.}, left). 
These patches are increasingly recognized as prominent mesoscale structures in the highly Alfv\'enic young solar wind \citep{Badman2026Review, Shi2022, Raouafi2023}. Mesoscale structures are intermediate structures larger than the small kinetic scale structures but smaller than the global large scale structures such as corotating interaction regions (CIRs), the heliospheric current sheets (HCS) or large coronal mass ejections (CMEs; \citealt{Viall_DeForest_Kepko_2021}). 

Farther from the Sun, the solar wind is also observed to often contain structures with mesoscale speed and density fluctuations in the solar wind that are associated with coherent large scale structure. These are typically referred to as "microstreams", \citep{Neugebauer1995, Horbury2023}. Microstreams were originally identified from Ulysses \citep{Wenzel1992} observations between 1-5AU as mesoscale velocity enhancements with amplitudes of approximately $40~\mathrm{km\,s^{-1}}$ from the running average speed, with higher proton temperatures, enhanced ion fluxes, and compression on their leading edges with rarefaction on their trailing edges (Figure \ref{Figure 1.}, right). More recently, \citet{Neugebauer_Sterling_2021} demonstrated that these velocity peaks are associated with patches of switchbacks.

In terms of longitudinal extent, SB patches and microstreams are similar in scale, and the arc-like SB structures are reminiscent of the background velocity enhancement characteristic of microstreams. This raises the hypothesis that microstreams are the evolved state of these near-Sun structures. Throughout this work, we test this hypothesis. Under this idea, the mesoscale structures should remain traceable with heliocentric distance even as their Alfv\'enic and plasma properties evolve. The proposed evolution of near-Sun SB patches into microstream like structures is illustrated in Figure \ref{Figure 1.}.

Determining whether these mesoscale structures advect coherently from the low corona to Earth and beyond is critical both for understanding the fundamental physics of the heliosphere and for establishing the background state of the solar wind against which space weather disturbances evolve and propagate.

An important question is whether space weather models should seek to capture these scales in order to provide accurate forecasts. SBs carry a significant amount of energy and may play a role in the heating and acceleration of the solar wind \citep{Rivera2024, Rivera2025}. \citet{Rivera2024} found that Alfvén/SB waves transferred a significant amount of energy and momentum into the solar wind as they evolved, providing a possible mechanism by which strongly Alfv\'enic SB patches near the Sun could contribute to the preservation of microstream-like velocity structures farther out, which might otherwise be expected to smoothen out by processes such as stream interactions. 

If these structures survive to 1 AU as microstreams, tracing them back to their source(s) informs on their formation \citep{Rivera2024b_patches,Coburn2026} and may preserve clues on the energization and heating related to their evolution. More generally, future heliophysics increasingly rely on coordinated multi-spacecraft observations, making robust methods for identifying corresponding plasma structures increasingly important.

Previous studies have investigated the proposed evolution of SB patches into microstreams. \citet{Soni2024} analyzed data from a radial alignment between {\it{Parker}} and SolO during {\it{Parker}}'s Encounter 9 (E9) and reported that the near-Sun parcel observed by {\it{Parker}} was a SB patch, with a background proton speed comparable to that of the surrounding pristine solar wind, whereas the farther out parcel observed by SolO exhibited microstream characteristics, with an enhanced background proton speed. The SolO microstream also contained, on average, approximately $30\%$ fewer switchbacks than the near-Sun SB patch. \citet{Soni2024} interpreted the reduction in switchback occurrence together with the enhanced background flow as evidence consistent with magnetic relaxation, in which relaxation of switchbacks may transfer energy to and accelerate the background plasma as the structure propagates outward. However, establishing an evolutionary connection requires identifying the same individual mesoscale structures at multiple heliocentric distances, and the compared intervals in \citet{Soni2024} were located on opposite sides of the HCS, limiting the confidence that they represented multi-point observations of the same plasma structure. 

Similarly, \citet{Horbury2023}, using a near-radial alignment between {\it{Parker}} and SolO, reported that the longitudinal scales of these mesoscale structures persisted while the prominent magnetic signatures of switchbacks were no longer clearly identifiable farther from the Sun. While their results provide evidence that switchback patches are related to microstreams, they do not establish a one-to-one correspondence between individual near-Sun SB patches and their farther out counterparts. 

Although these studies have demonstrated similarities and possible evolutionary connections between SB patches and microstreams, a well-constrained one-to-one mapping of individual patches from the near-Sun solar wind to 1 AU has not been established.

In this work, we connect SB patches and microstream mesoscale structures through detailed mapping of individual patches from the near-Sun solar wind out to 1 AU. Specifically, we investigate how the plasma and magnetic properties of these structures evolve with heliocentric distance and whether their evolution is consistent with the proposed SB patch to microstream hypothesis. 
By flying much closer to the Sun than any previous mission, now reaching below 10 $R_{\odot}$, {\emph{Parker}} provides an opportunity to observe SB patches coherently in less evolved solar wind.
We track the SB patches as they evolve with growing heliocentric distance through an alignment between {\it{Parker}}, SolO, and \textit{Wind} \citep{ogilvie_1995, 22_Lepping1995} during {\it{Parker}}’s Encounter 19 (E19). This alignment enables us to observe the same solar wind streams at three heliographic distances, with minimal longitudinal separation, covering solar distances from 11 to 210 $R_{\odot}$. All three spacecraft observed the same coherent, long lived solar wind streams that emanated from a large and well observed coronal hole source on the Sun \citep{Rivera2026_eclipse}. 

The solar wind stream is bounded by two heliospheric current sheets, making it easily identifiable between spacecraft. Due to the well-defined nature of the solar wind stream across large distances, we employ Dynamic Time Warping (DTW)--an algorithm that allows for stretching and compression of time series to align similar structures \citep{DTW}, as the primary mapping technique. DTW has been applied productively in different areas of solar physics research, such as in calibration of Faraday cups \citep{Ahmed_L_2025} and for space weather forecast evaluation \citep{Samara2022}. Unlike ballistic mapping, the most common approach \citep[e.g.][]{Nolte1973,Badman2020} which assumes no stream-stream interactions and allows streamlines to cross each other, DTW preserves ordering while allowing corresponding structures to undergo longitudinal stretching and compression. This makes DTW particularly useful for the present analysis, where the objective is to identify the same evolving structures. We develop a Physics-Informed Dynamic Time Warping (Pi-DTW) approach that incorporates physical conservation constraints in the DTW cost function while allowing longitudinal stretching and compression of the mapped structures. This approach allows us to test whether the same sub-structures can be traced from {\it{Parker}} to 1 AU, how their plasma and magnetic properties evolve with heliocentric distance, and whether those properties at 1 AU are consistent with an evolution to microstreams.

This paper is organized as follows: Section~\ref{sec:data} describes the observations/data used in this study, and the unique E19 radial alignment.
Section~\ref{sec:Mapping Methodology} describes the data processing, initial patch identification, mass  and energy flux comparisons, and the DTW/Pi-DTW mapping methodology.
Section~\ref{sec:Results and Discussions} presents the resulting multi-spacecraft mappings and discusses the radial evolution of the mapped structures and comparison with the observational characteristics of microstreams.
Lastly, Section~\ref{sec:Summary and Conclusions} summarizes the main conclusions, and discusses the implications for the radial evolution of switchback patches, the relationship between SB patches and microstreams, the broader applicability of Pi-DTW to multi-spacecraft heliophysics, and the future directions of this work.

\section{Observational Datasets} 
\label{sec:data}

\subsection{Spacecraft Data Sets}
We analyze magnetic field and plasma measurements from {\it{Parker}}, SolO, and \textit{Wind}, all accessed with \texttt{PySPEDAS} \citep{pySPEDAS}. Where available, we use magnetic field vector measurements and proton, electron and alpha plasma properties (density, velocity and temperature), which allow us to compute or construct all relevant physical quantities for characterizing switchback patches and microstreams.
The instruments, data types, and variables used in this study are listed in Table \ref{tab:Table1}.

\subsubsection{Parker Solar Probe observations}

{\it{Parker}} magnetic field observations are provided by the FIELDS instrument suite \citep{Bale2016}.
Proton velocities and temperatures are obtained from the Solar Probe Analyzer-ions (SPAN-i; \citealt{Livi2022}) instrument of the Solar Wind Electrons Alphas and Protons (SWEAP) instrument suite \citep{Kasper2016}. 

Electron measurements were obtained from the Solar Probe Analyzer-Electrons (SPAN-e; \citealt{Whittlesey2020}) instrument of the SWEAP instrument suite, using the methodology of \citet{Halekas2020}. 

We use the electron density obtained from the quasithermal-noise (QTN) data product the FIELDS Radio Frequency Spectrometer (RFS; \citealt{Pulupa2017}) low-frequency range (LFR) spectra, following the method described by \citet{Moncuquet2020}. Here, we approximate $N_e = N_p$ (density for electrons and protons respectively) to estimate the proton density, which introduces an additional source of uncertainty. In the presence of alpha particles, $N_p<N_e$, but at the same time, the alpha particle density (derived from SPAN-i) is a partial moment i.e an underestimate of the true alpha particle density. These two effects can compensate each other, so it is not clear whether the derived quantities are over or underestimated. 

Alpha particle density and velocity measurements used for the helium abundance and alpha-proton differential streaming calculations were obtained from SPAN-i moments data, after applying SPAN-i field of view (FOV) filters to exclude measurements with significant FOV aberration of the particle distribution from Alfv\'enic velocity spikes.

\subsubsection{Solar Orbiter observations}

SolO magnetic field measurements were obtained from the MAG fluxgate magnetometer \citep{Horbury2020}. Plasma measurements were obtained from the Proton-Alpha Sensor (PAS), part of the Solar Wind Analyzer (SWA) instrument suite \citep{Owen2020}. 
Alpha particle or electron measurements were not available for SolO; consequently, neither SolO alpha abundance nor alpha-proton differential streaming or electron measurements are reported.

\subsubsection{\textit{Wind} Observations}
\textit{Wind} magnetic field measurements were obtained from the Magnetic Field Investigation (MFI; \citealt{Lepping1995_MFI}), and proton plasma measurements as well as alpha density and speed measurements were obtained from the Solar Wind Experiment (SWE; \citealt{Ogilvie1995}) Faraday Cup of the \textit{Wind} spacecraft \citep{Wilson2021Wind}. 
The data source for magnetic field vector components is \texttt{\detokenize{WI_H0_MFI}} while for \textit{Wind} SWE it is \texttt{\detokenize{WI_H1_SWE}} where the plasma variables in it are: proton thermal speed \texttt{\detokenize{Proton_W_nonlin}}, speed \texttt{\detokenize{Proton_V_nonlin}}, and density \texttt{\detokenize{Proton_Np_nonlin}}.

Since SB patches and microstreams have characteristic longitudinal scales, we principally assess measurements as a function of Carrington longitude, using the spacecraft trajectories to convert from space to time. The plasma and magnetic field time series were median filtered by being grouped into $1/21${\textdegree}  wide bins and taking the median value within each bin to reduce spikes, and then resampled these filtered data onto the same 252 point longitude grid, with longitude in units of degrees. The $1/21${\textdegree} window was chosen after testing several bin widths and identifying the value that provided the most effective reduction of isolated spikes while preserving the overall "arc-like" structure of the observations required to identify SB patches and microstreams.

\startlongtable
\begin{deluxetable*}{l l l}

\tablecaption{
\label{tab:Table1}
Data sources and primary variables used in this study.
}

\tablehead{
\colhead{Spacecraft / Instrument} &
\colhead{Data Type} &
\colhead{Variables}
}

\startdata
Parker/FIELDS & Magnetic field & $B_R$, $B_T$, $B_N$, $|B|$\\
\tableline
Parker/SPAN-i & Plasma data and Helium & $V_p$, $T_p$, $n_\alpha$, $V_\alpha$\\
\tableline
Parker/SPAN-e & Electron data & $n_e$, $T_e$ \\
\tableline
Parker/LFR QTN & Plasma density & $n_p$\\
\tableline
SolO/MAG & Magnetic field & $B_R$, $B_T$, $B_N$, $|B|$\\
\tableline
SolO/SWA/PAS & Plasma data & $n_p$, $V_p$, $T_p$\\
\tableline
\textit{Wind}/MFI & Magnetic field & $B_X$, $B_Y$, $B_Z$, $|B|$\\
\tableline
\textit{Wind}/SWE & Plasma data and Helium &
\shortstack[l]{
  \detokenize{Proton_Np_nonlin}, \detokenize{Proton_V_nonlin},\\
  \detokenize{Proton_W_nonlin}, \detokenize{Alpha_Na_nonlin},\\
  \detokenize{Alpha_V_nonlin}
}\\
\enddata
\end{deluxetable*}

\subsection{{\it{Parker}} Encounter 19 timeline}

Here, we review the spacecraft orbital geometry which allows us to track the detailed substructure of a solar wind stream from 11 $R_{\odot}$ to 1~AU. 

\begin{figure*}[ht!]
\centering
\includegraphics[width=0.42\linewidth]{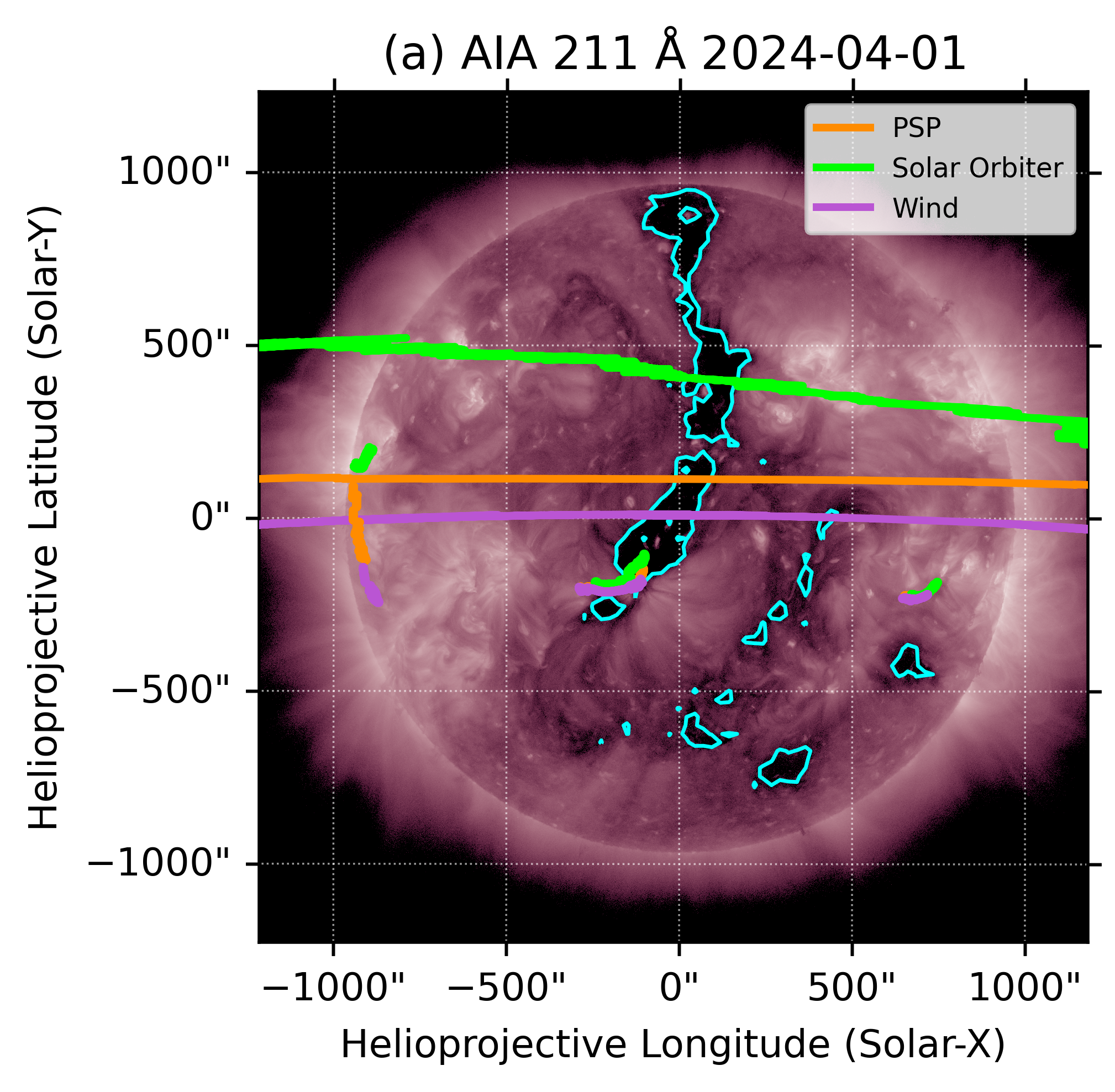}
\hfill{}
\includegraphics[width=0.55\linewidth]{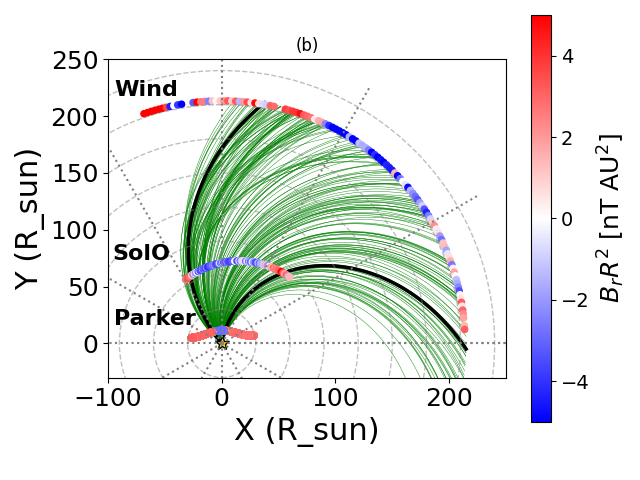}
\caption{(a): Coronal hole source region and ballistically mapped spacecraft trajectories. An SDO/AIA 211~\AA{} image of the Sun shows the narrow negative polarity coronal hole sourcing the streams studied in this work. Orange, lime, and magenta curves show the ballistically projected trajectories of {\it{Parker}}, SolO, and \textit{Wind}, respectively, at $2.5\,R_\odot$. 
Each stream's footpoints are indicated in matching colors.
While the trajectories cross the solar disk at different latitudes, the corresponding footpoints all map to the coronal hole's southern tip.
(b): Orbital trajectories during all three spacecraft traversals, shown in Cartesian coordinates. Trajectories are colored by the radial magnetic field scaled as $B_RR^2$ where the $R$ in this expression is in units of $AU$. Green curves show ballistic streamlines calculated using the measured SolO solar wind speed, while the black curves highlight streamlines corresponding to the source surface longitudes of the current sheet crossings that bound the interval analyzed in this study.}
\label{Figure 2.}
\end{figure*}

{\it{Parker}} Encounter 19 spanned March 25-April 4, 2024, with {\it{Parker}} coming within $11R_{\odot}$ (7.26 million kilometers) of the Sun's surface. During this time, \emph{Parker} was in conjunction with the Earth and SolO within a solar wind stream emerging from a large negative-polarity transequatorial coronal hole, as shown in Figure~\ref{Figure 2.} and further discussed by \citet{Rivera2026_eclipse}.

Figure \ref{Figure 2.}a shows an extreme ultraviolet image of the Sun at $211$ {\AA} obtained by the Atmospheric Imaging Assembly (AIA; \citealt{Lemen2012}) on board the Solar Dynamics Observatory (SDO), with ballistically mapped spacecraft footpoints at the source surface ($2.5R_{\odot}$) for each spacecraft during the alignment period. The coronal hole source region is outlined in cyan. Coupled to a Potential field Source surface (PFSS; \citealt{Altschuler1969_pfss, Schatten1969_pfss} coronal model, as implemented in \verb+sunkit-magex+ \citealt{Stansby2020}), we trace the source surface trajectories to their photospheric footpoints which are color coded to correspond to each trajectory.
The spacecraft footpoints fall within the same coronal hole region and overlap most closely on the left hand side, while the SolO footpoints become separated from \emph{Parker} and \textit{Wind} toward the right hand side, further discussed in the context of the SB patches in Section \ref{sec:Summary and Conclusions}. This alignment enabled multi-point sampling of the same solar wind stream originating from the same coronal-hole region. 

Figure \ref{Figure 2.} (b) shows the orbital trajectories in the solar equatorial plane over heliocentric distances of 11-210 $R_{\odot}$ color coded with the radially scaled magnetic field in the radial direction ($B_RR^{2}$) as the color map, where red is for positive polarity (away from the Sun) and blue is for negative polarity (towards the Sun). Regions where the polarity changes represent crossings of the heliospheric current sheet (HCS).
The green spiral lines represent ballistic streamlines/flow line that project a given point on the spacecraft trajectory back to the source surface \citep{Nolte1973, Badman2020}. The black spiral lines represent the ballistic flow lines seeded from the HCS crossings measured at {\it{Parker}} and illustrate the limits of ballistic mapping: while the mapping from {\it{Parker}} to SolO accurately connects the current sheets, it is much less accurate a prediction at 1AU. As shown in the figure, the stream and both current sheets were measured at least three times as shown directly in the measured magnetic polarity. 

During the E19 period, we therefore identify the solar wind stream at each spacecraft using the two HCS crossings as markers, avoiding needing to rely on ballistic mapping for the analysis presented in this work. {\it{Parker}} traversed the current sheet bounded stream (marked by magnetic field polarity reversals in Figure \ref{Figure 2.}b) at a heliocentric distance of approximately 11 to 90 $R_{\odot}$ from the Sun's surface. SolO subsequently sampled the corresponding stream from April 7-13, 2024 at approximately 50 to 60 $R_{\odot}$, while \textit{Wind} sampled the stream near L1 from April 4-8, 2024 at approximately $1$~AU.

\section{Mapping Methodology} 
\label{sec:Mapping Methodology}

Once the solar wind stream intercepted by each spacecraft was identified, the methodology for this study follows four main steps: (1) Data pre-processing and identifying SB patches, (2) mapping corresponding structures between spacecraft using Dynamic Time Warping (DTW), and (3) assessing validity of the mapping from conserved quantities and (4) using these mappings to investigate the evolution of these structures from \emph{Parker} outwards to determine whether SB patches evolve into microstreams.


\subsection{Data pre-processing and Initial patch identification}

High-cadence magnetic field and plasma observations from {\it{Parker}}, SolO, and \textit{Wind} are pre-processed using standard data cleaning procedures such as removing NaNs and median filtering. 

\begin{figure*}[ht!] 
\centering
\includegraphics[width=1\textwidth]{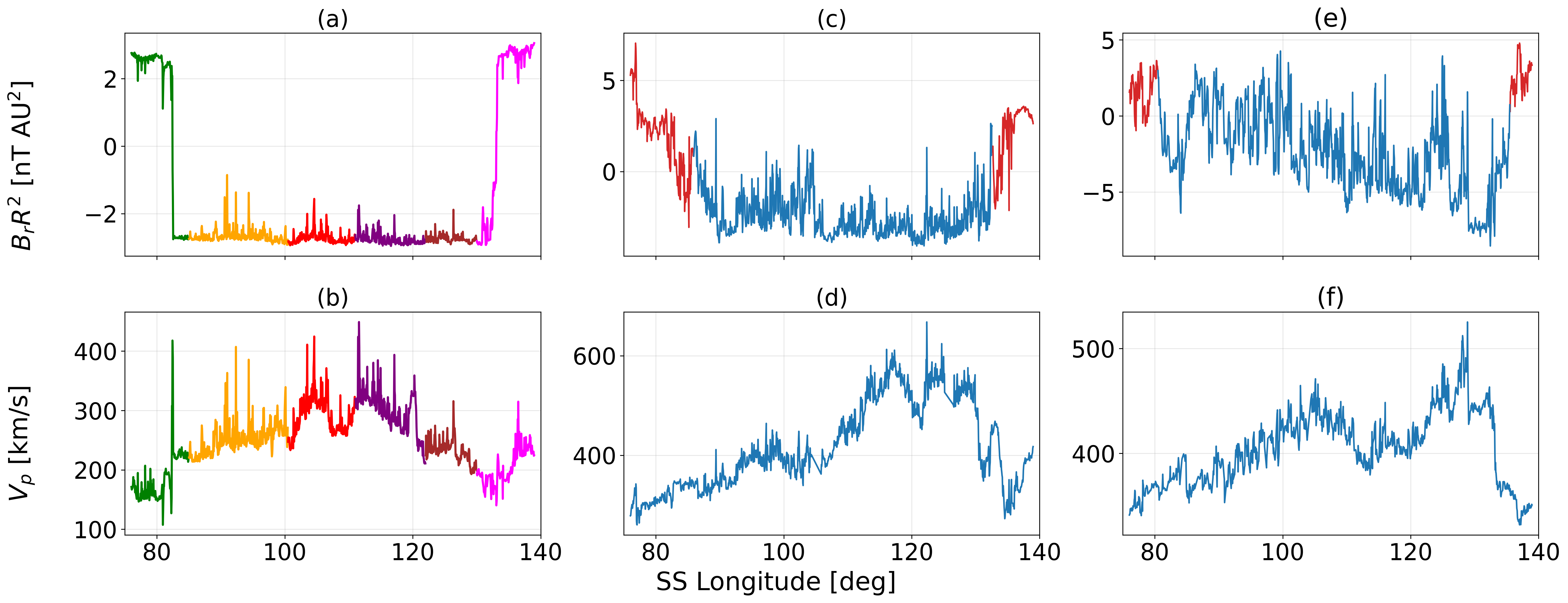}
\caption{Initial identification of SB patches and current sheet crossings in source surface longitude. The top panels show the scaled radial magnetic field, $B_RR^2$, and the bottom panels show the proton radial velocity, $V_p$, for {\it{Parker}} (left column), SolO (middle), and \textit{Wind} (right). The {\it{Parker}} measurements correspond to the March 29-30, 2024 perihelion period, while the SolO measurements span April 7-13, 2024, and \textit{Wind} measurements span April 4-8, 2024. Colors in {\it{Parker}} plots identify the initially selected patches and CS crossings based on visual identification. Transition from red to blue colors in SolO and \textit{Wind} mark polarity reversal or CS crossings. }
\label{Figure 3.}
\end{figure*}

Candidate SB patches were first visually identified by inspecting the magnetic field and plasma data. A reference set of patches is defined and fixed at {\it{Parker}} (color coding in Fig \ref{Figure 3.}, left hand column). As will be discussed below, the corresponding identification at SolO and L1 was subsequently revised and refined using conserved physical quantities, such as mass flux and total energy flux and combined with the DTW technique to produce a physics-informed automated procedure. The following steps describe how the initial boundaries were estimated and tested.

\subsubsection{Ballistic Mapping and Current Sheet Boundaries}

As mentioned above, the limitation of ballistic mapping is illustrated in Figure~\ref{Figure 2.}b, the ballistic streamlines (black) do not intersect the observed current sheet crossings at their exact locations and flows lines (green) intercept. This discrepancy reflects the simplifying assumption of ballistic propagation as an idealized Parker spiral. Namely, this approach does not account for interactions between neighboring streams, allowing the flow lines to unphysically overlap. It also does not account for any acceleration, since a constant speed is assumed to create the flow line. All of these contribute to the uncertainty in the mapping between solar wind measured by different spacecraft at different heliocentric positions. While this technique is sufficient to identify and relate the two bounding current sheets, and therefore identify the large scale stream to be investigated at each measurement point, with our current goal of mapping and tracking mesoscale substructure outwards, we seek higher fidelity mapping. These limitations further motivate the use of more robust mapping methods.

In light of this, we first reproject the spacecraft positions to 2.5 $R_\odot$ using ballistic mapping with a \textit{uniform speed} across all data. This performs an approximate initial alignment between the current sheet crossings observed by each spacecraft and aligns them in longitude, but keeps the ordering of data points in longitude preserved.

\subsubsection{Visual Patch Identification}

Figure \ref{Figure 3.} presents a summary of the principal quantities ($B_R$ and $V_p$) used for identifying SB patches and microstreams, as measured in source surface longitude (produced using the uniform-speed deprojection described above) across the stream of interest at each measurement point ({\it{Parker}}, SolO, L1 : left, middle and right hand columns respectively). 
The colors in {\it{Parker}} [panels (a) and (b)] identify the initially selected patches and HCS crossings based on visual identification. 
Panels (c),(d),(e),(f) are plotted for a comparison of the same quantities that correspond to SolO and \textit{Wind}. The longitudes colored in green and magenta for {\it{Parker}} are locations of HCS crossings while the longitudes colored in red for SolO and \textit{Wind} are locations of HCS crossings.

Previous studies have commonly identified SB patches through visual inspection of magnetic field and velocity signatures \citep[e.g.][]{Bale_2021,Fargette_2021,Shi2022}. We therefore used visual identification 1) to identify SB patches at {\it{Parker}} which are quite unambiguous, and 2) as an initial baseline for the mapping further out.

During the initial phase, a team visual identification exercise was performed where six users (first author and five other co-authors) individually tried to classify the similarity by eye using a python interactive plotter. The interactive plotter synchronously color-codes the same regions in both magnetic field and speed plots for the solar wind within the bounded HCS crossings, providing the user with an additional patch identification constraint. 

In the {\it{Parker}} observations, candidate SB patches were identified as regions containing arc like structures in magnetic field strength and velocity that are accompanied by spikes, which are also correlated with enhancements in $V_p$(proton speed) as can be seen in Figure \ref{Figure 3.}, first column. These structures are 5-10 degrees in longitude. 

All six users independently identified four patches at each spacecraft and a consensus was reached. These estimated patch locations for {\it{Parker}} are plotted in the first column of Figure \ref{Figure 3.} marked by orange, red, purple, and brown colors bounded by the left side HCS crossing (marked by green) and the right side HCS crossing (marked by magenta). This color scheme will be followed to track the mapping as our technique is improved in the subsequent sections. Thus, the users agreed on the number and ordering of the structures and were nearly perfectly in agreement at {\it{Parker}}. The SB patches identified at {\it{Parker}} were therefore fixed at this stage. However, for the further out measurements (SolO: \ref{Figure 3.} Panels (c,d) and \textit{Wind}: Panels (e,f)) most users selected slightly different estimates for the location of each patch, and so other than the location of the HCS boundaries, we do not colorize the patches/microstream candidates at this stage.
This variability between users demonstrated that the patch boundaries from visual identification need to be refined.  Physical conservation laws were therefore used to further constrain their locations.

\subsubsection{Mass- and Energy-Flux Constraints} \label{subsec:mass_energy_constraints}

The next step was using the overall mass and magnetic flux conservation to confirm the feasibility of SB patch mapping. The mass flux (as it is mass flux per unit solid area i.e. the flux of plasma particles in a "flux tube"), is expected to remain consistent within a radially or spherically symmetric expanding flux tube. More generally, the magnetic flux ($B_R R^2$) tracks deviations from this perfect spherical expansion picture with the quantity $g = nv/B_R$ expected to be conserved in general. Therefore, we can compare these quantities for coinciding SB patches at each heliocentric location for consistency. 

\begin{figure*}[ht!]
\centering
\includegraphics[width=0.49\textwidth]
{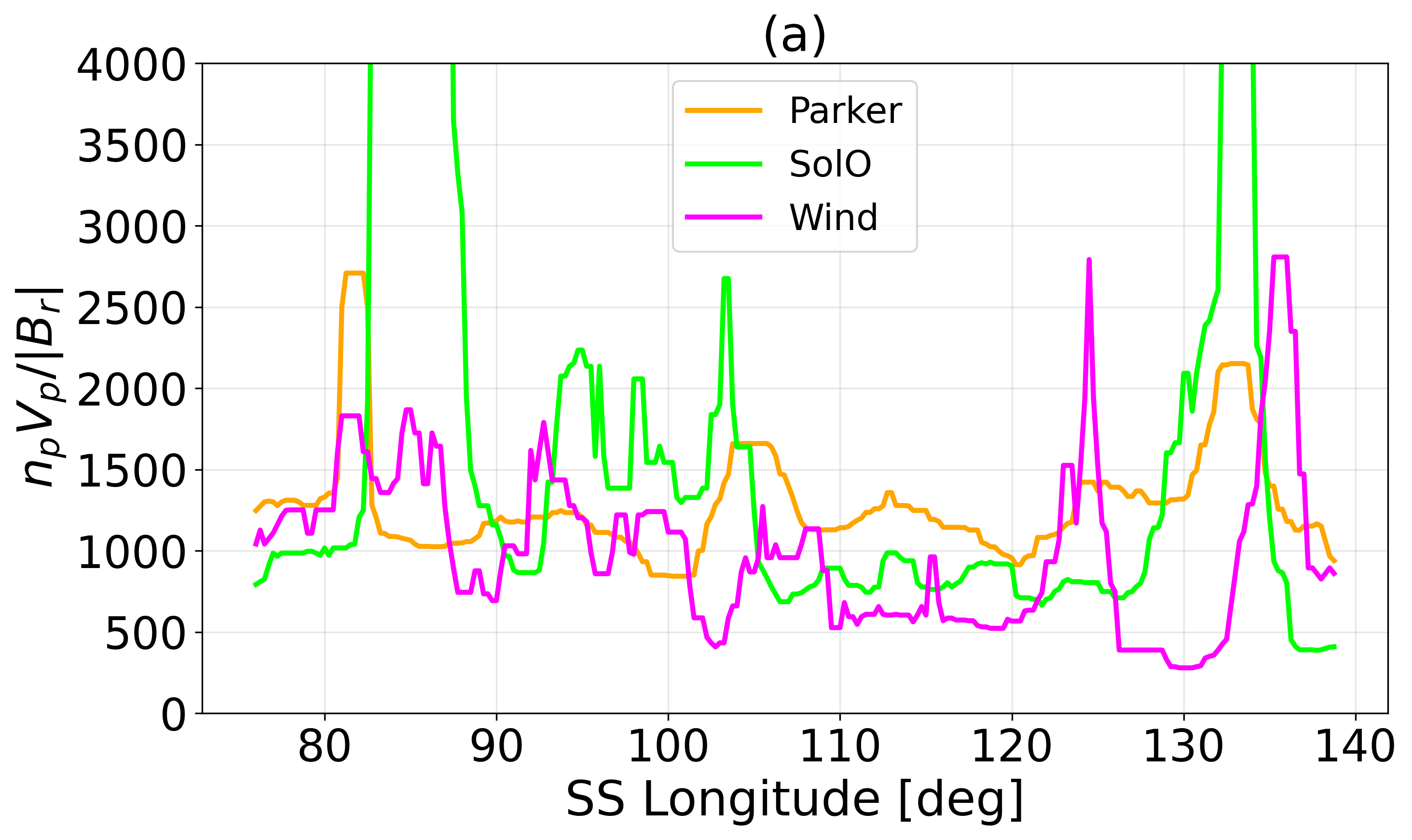}
\hfill
\includegraphics[width=0.49\textwidth] 
{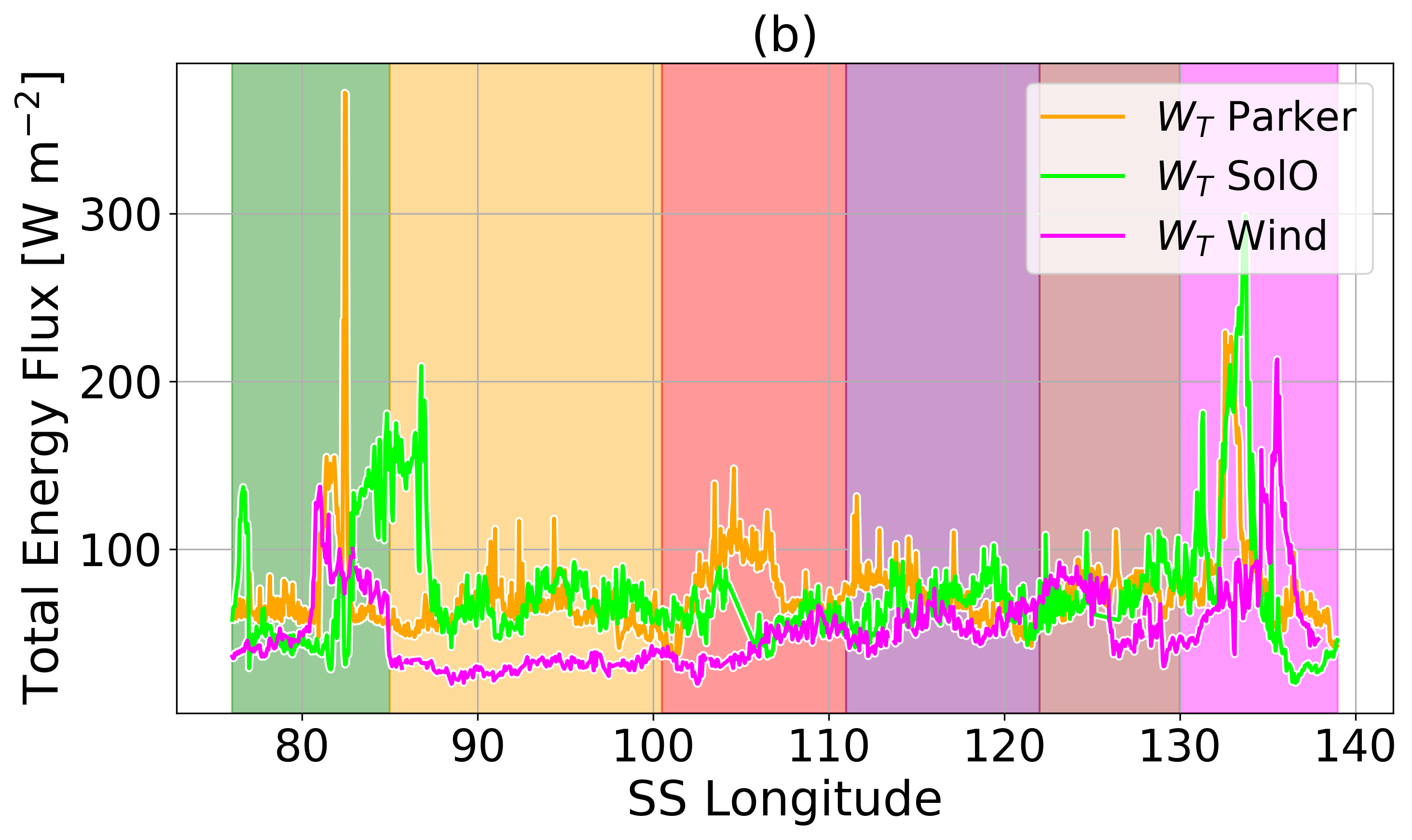} 
\caption{Comparison of mass and energy flux across the mapped solar wind patches observed by {\it{Parker}}, SolO and \textit{Wind}. (a)Ratio of mass flux to magnetix flux, $g = nv/B_R$ as a function of source surface longitude. (b) Total energy flux as a function of source surface longitude for {\it{Parker}}, SolO, and \textit{Wind}. The colored shaded regions identify the solar wind patches defined by the mass flux based refined boundaries for {\it{Parker}}. Spacecraft are distinguished by line color (orange for {\it{Parker}}, lime for SolO, and magenta for \textit{Wind})
}
\label{Figure 4.}
\end{figure*}

First we examine the value of $g$ over the entire stream (all patches), plotted as a function of Source Surface (SS) Longitude, shown in Figure \ref{Figure 4.}a. We observe approximate conservative behavior in between the two current sheets at each distance at a given longitude and with all spacecraft showing a weak decrease in $g$ from left to right. The mass and magnetic flux at each measurement point are shown independently in Appendix Figure \ref{fig:appendix_npVpBr} where it is observed that mass and magnetic flux are conserved independently from {\it{Parker}} to SolO, while at \textit{Wind} there appears to be evidence of non-radial flux tube expansion where the left side carries much weaker magnetic flux and correspondingly lower mass flux, however the stream is still well behaved because their ratio ($g$) remains close to consistent with the other measurement points. Overall, this is a sanity check to motivate that, while we expect the substructure to evolve, the same stream as a whole is captured at each measurement location.

Later, we will utilize the radially scaled proton mass flux used in the longitude series comparison on a patch by patch basis, which was calculated as,

\begin{equation}
M_F
=
n_pV_pR^2.
\end{equation}

while the corresponding mass flux through the mapped solid angle was calculated as,

\begin{equation}
\dot{M}
=
n_pV_pr^2\lambda.
\end{equation}
And the cumulative mass flux as
\begin{equation}
M_{F,\mathrm{cum}}(\lambda)
=
\int_{\lambda_{\rm start}}^{\lambda_{\rm end}}
M_F\,d\lambda'.
\end{equation}

This integral will be evaluated as a cumulative sum of $M_F\,\Delta\lambda$ over the longitude samples, where the integral bounds are the start and end longitudes of the individual patches.

The Energy Flux components of the solar wind were also calculated for the entire window between the HCS crossings at each spacecraft traversal, following the energy flux terms from \citet{Rivera2024, Rivera2026_eclipse}: 

\begin{equation}
\begin{split}
    W_\text{Kinetic} = \frac{1}{2} m_jn_jV_{p}^3 \quad\\
    W_\text{Enthalpy} = \frac{\gamma}{\gamma-1} n_jk_BT_jV_{p} \quad\\
    W_\text{Alfvén~Wave} = V_{p} \left(\frac{3}{2}+\frac{V_A}{V_{\rm cm}}\right)\frac{\delta B^2}{2\mu_0}\\
    W_\text{Gravitational} = m_jn_jGM_{\odot}/R_{\odot}(1-R_{\odot}/r)V_p \quad.\\
    \end{split}
    \label{eq:energy_flux}
\end{equation}



\vspace{0.6cm}
The four terms represent the kinetic, enthalpy, Alfv\'en wave, and gravitational energy flux contributions, respectively. All the variables are listed in Table \ref{tab:Table1}. We use $\gamma = 5/3$. $G$ is the gravitational constant, and $M_{\odot}, R_{\odot}$ are the Sun's mass and radius, respectively. $k_B$~is the Boltzmann constant. Here, the subscript $j$ denotes the plasma species and the kinetic, enthalpy, and gravitational energy flux terms are calculated by summing the contributions of the species. For {\it{Parker}}, the kinetic, enthalpy, and gravitational energy density terms include contributions from protons, alpha particles, and electrons. For SolO, these terms were calculated using protons and where a helium contribution was required, a helium abundance of $4\%$ was assumed which corresponds to the average measured abundance at \emph{Parker} and \emph{Wind}. For \textit{Wind}, both protons and alpha particles were measured and included. At both SolO and \textit{Wind}, the electrons were assumed to be negligible for the kinetic and gravitational energy fluxes while for the enthalpy we assume the same proton to electron temperature ratio as found at \emph{Parker} as an estimate to its contribution. The Alfv\'en wave term is calculated separately using the magnetic field fluctuations.

\begin{figure*}[ht!]
\centering

\includegraphics[width=0.34\textwidth, trim=20 0 20 10, clip]{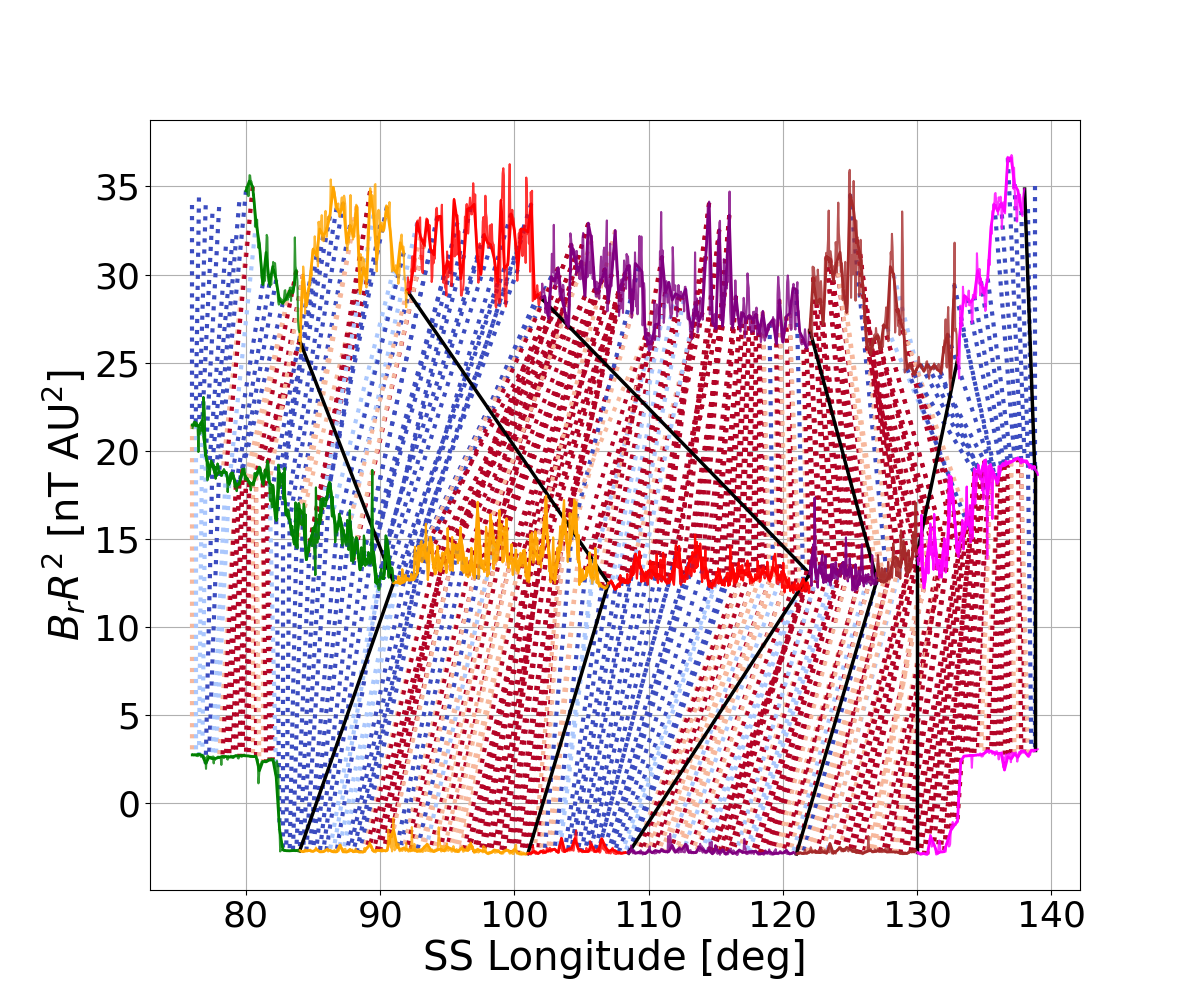}%
\includegraphics[width=0.34\textwidth, trim=20 0 20 10, clip]{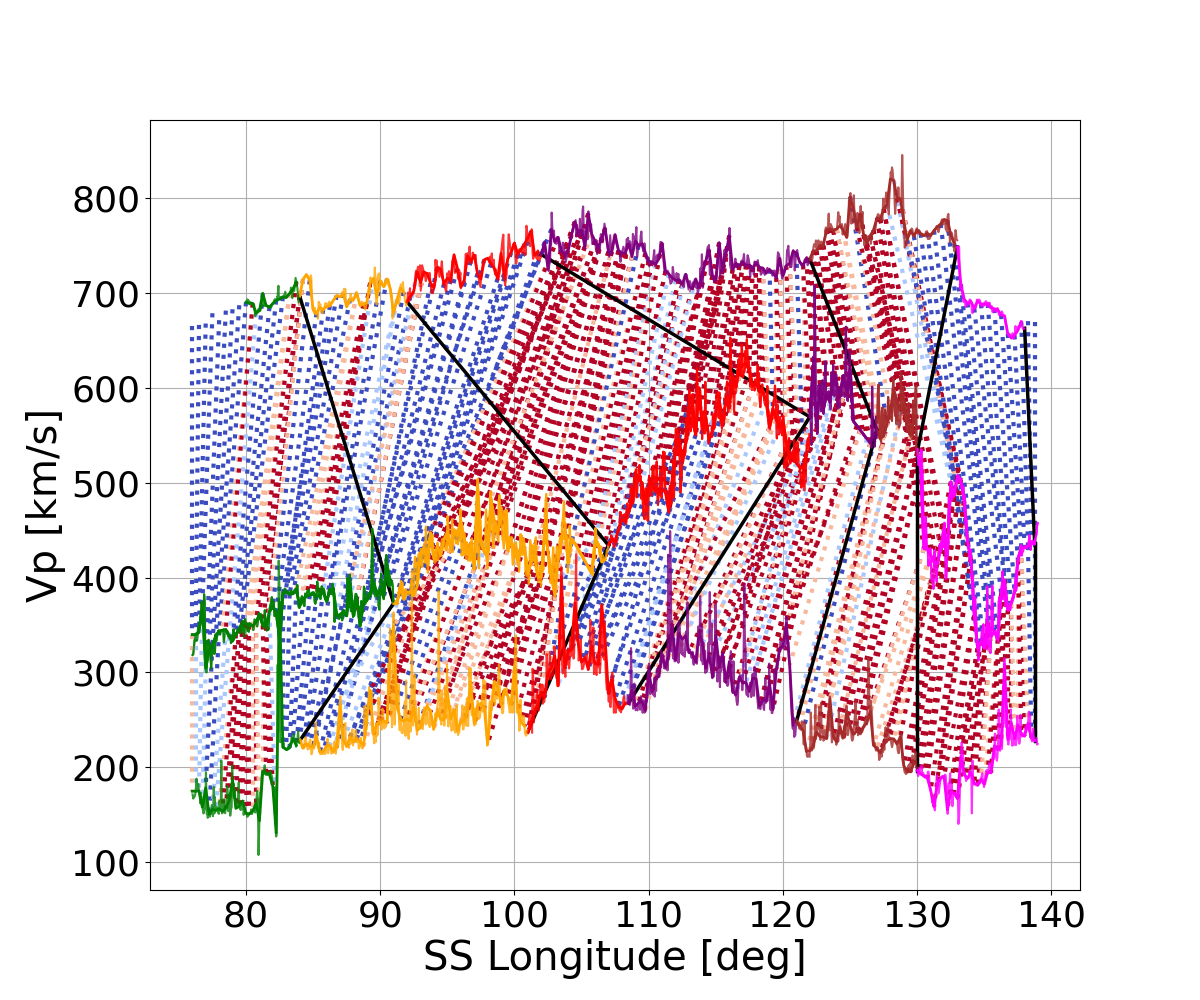}%
\includegraphics[width=0.32\textwidth, trim=0 0 0 0, clip]{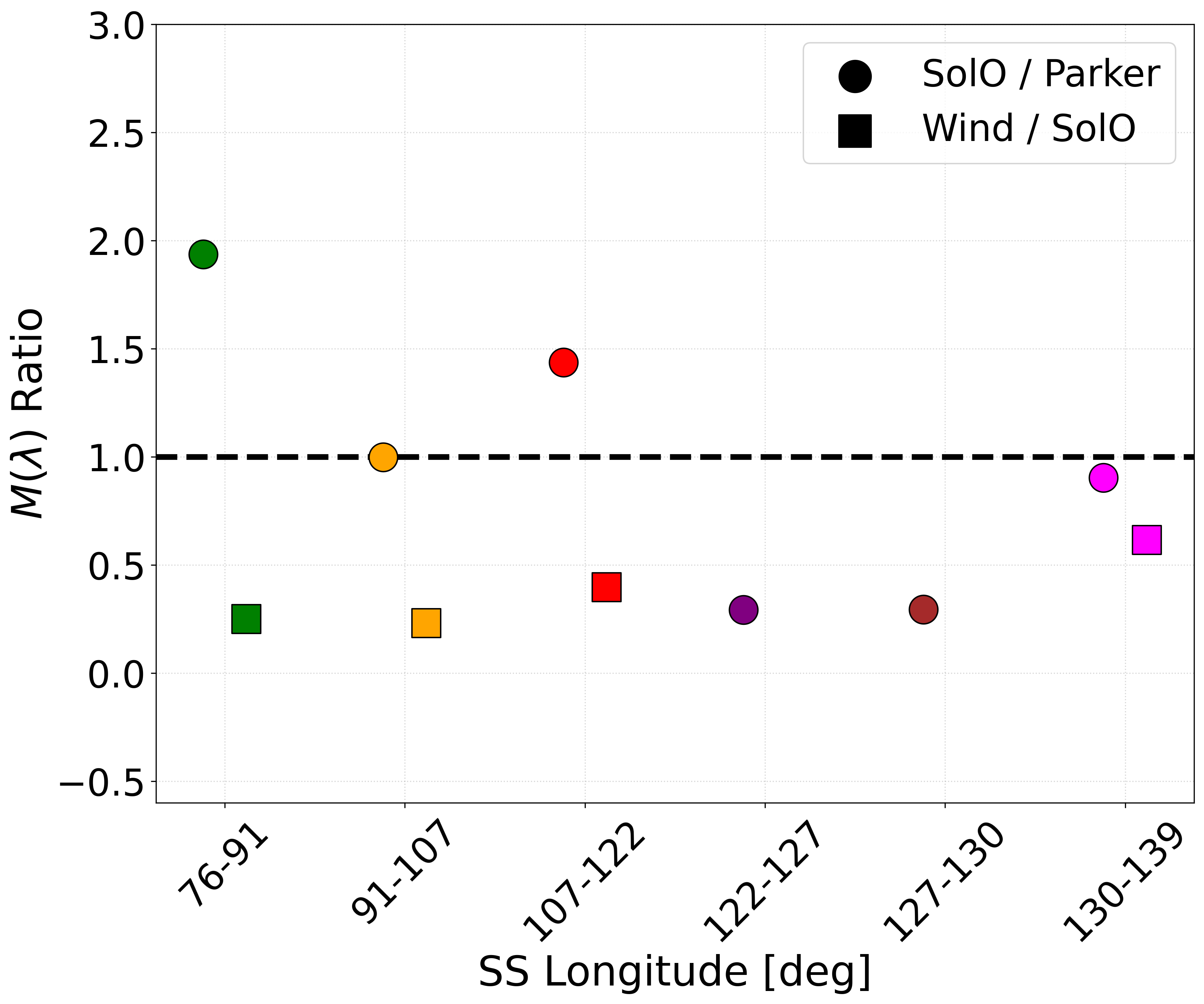}

\vspace{0.1cm}

\includegraphics[width=0.34\textwidth, trim=20 0 20 10, clip]{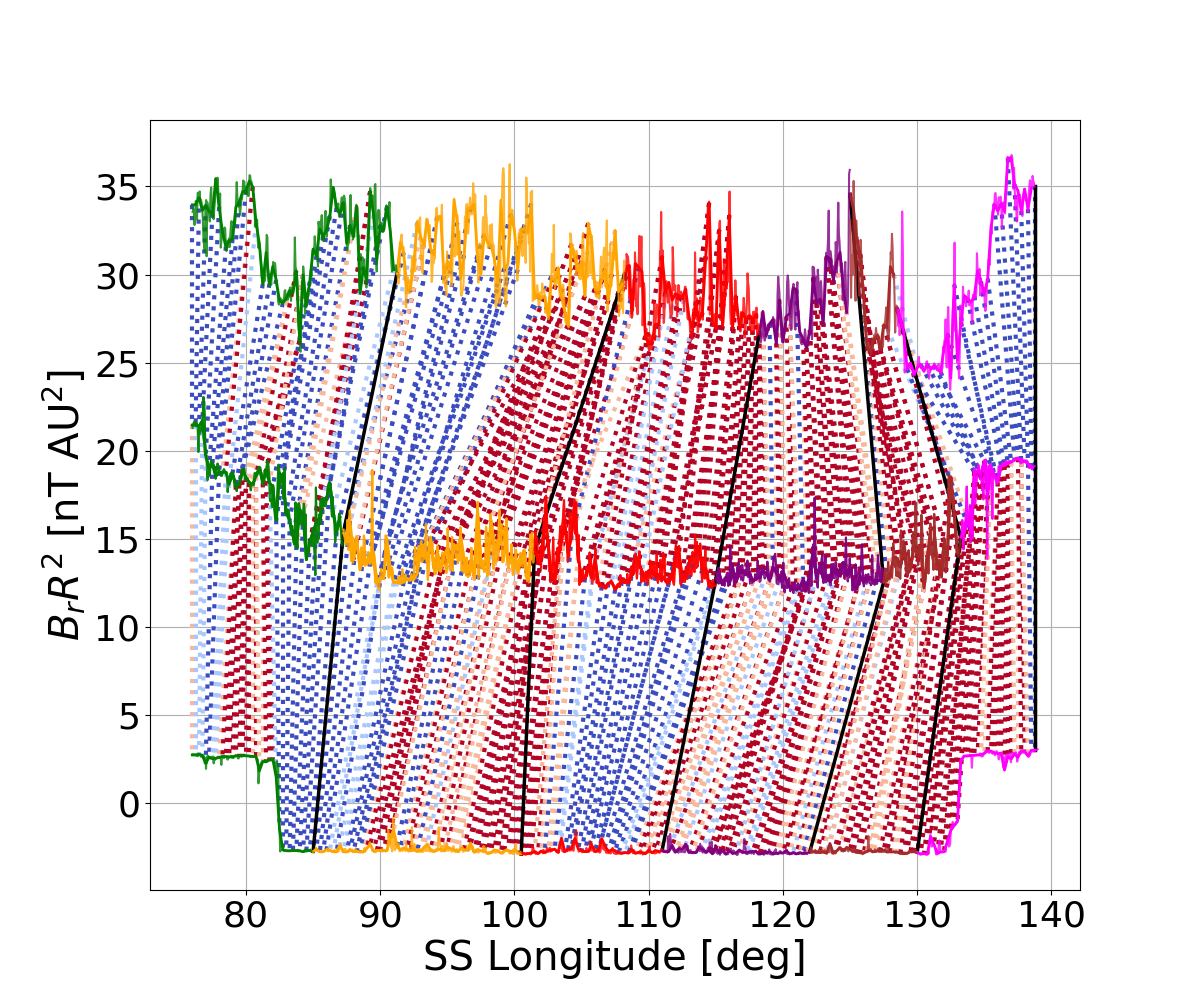}%
\includegraphics[width=0.34\textwidth, trim=20 0 20 10, clip]{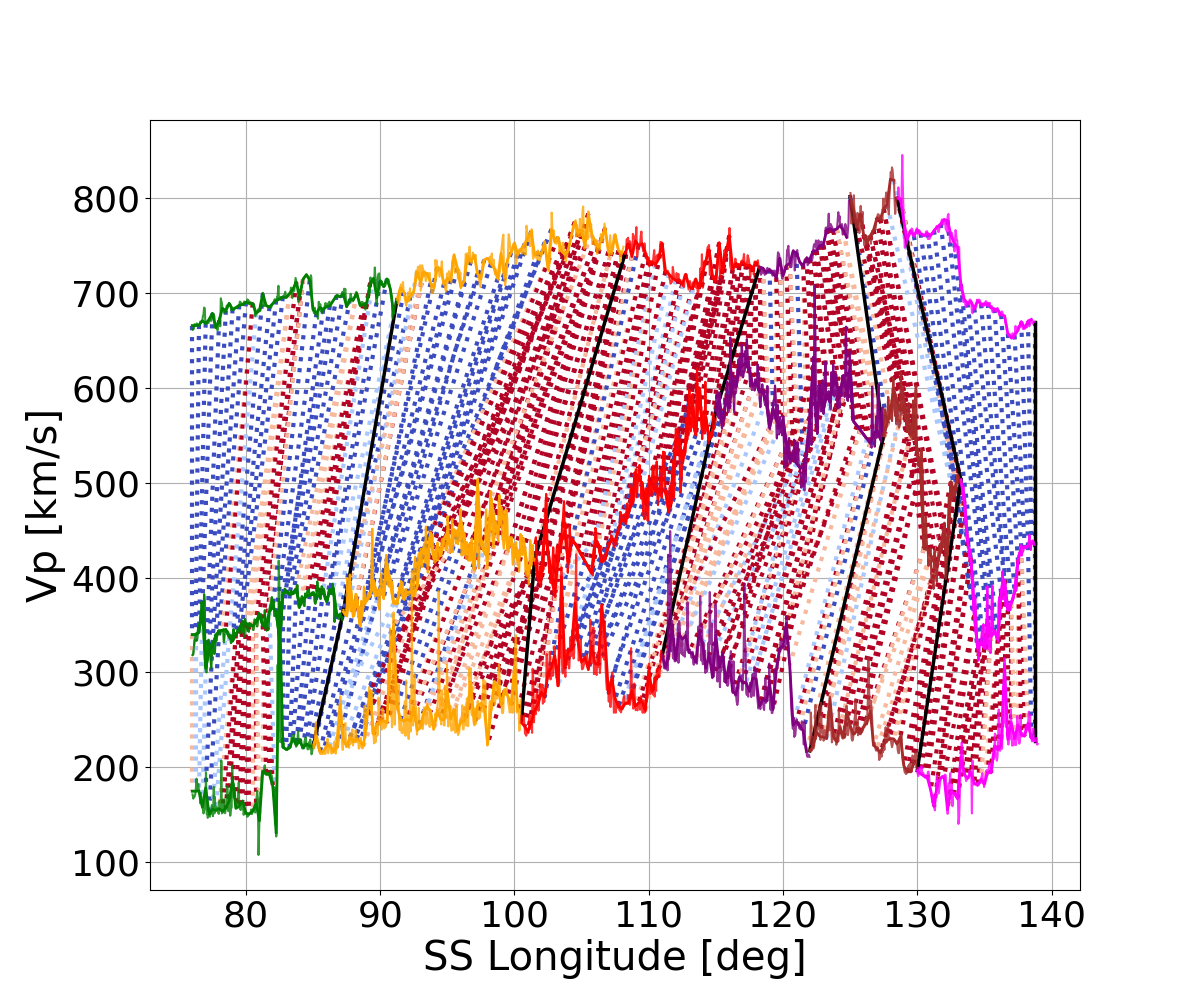}%
\includegraphics[width=0.32\textwidth, trim=0 0 0 0, clip]{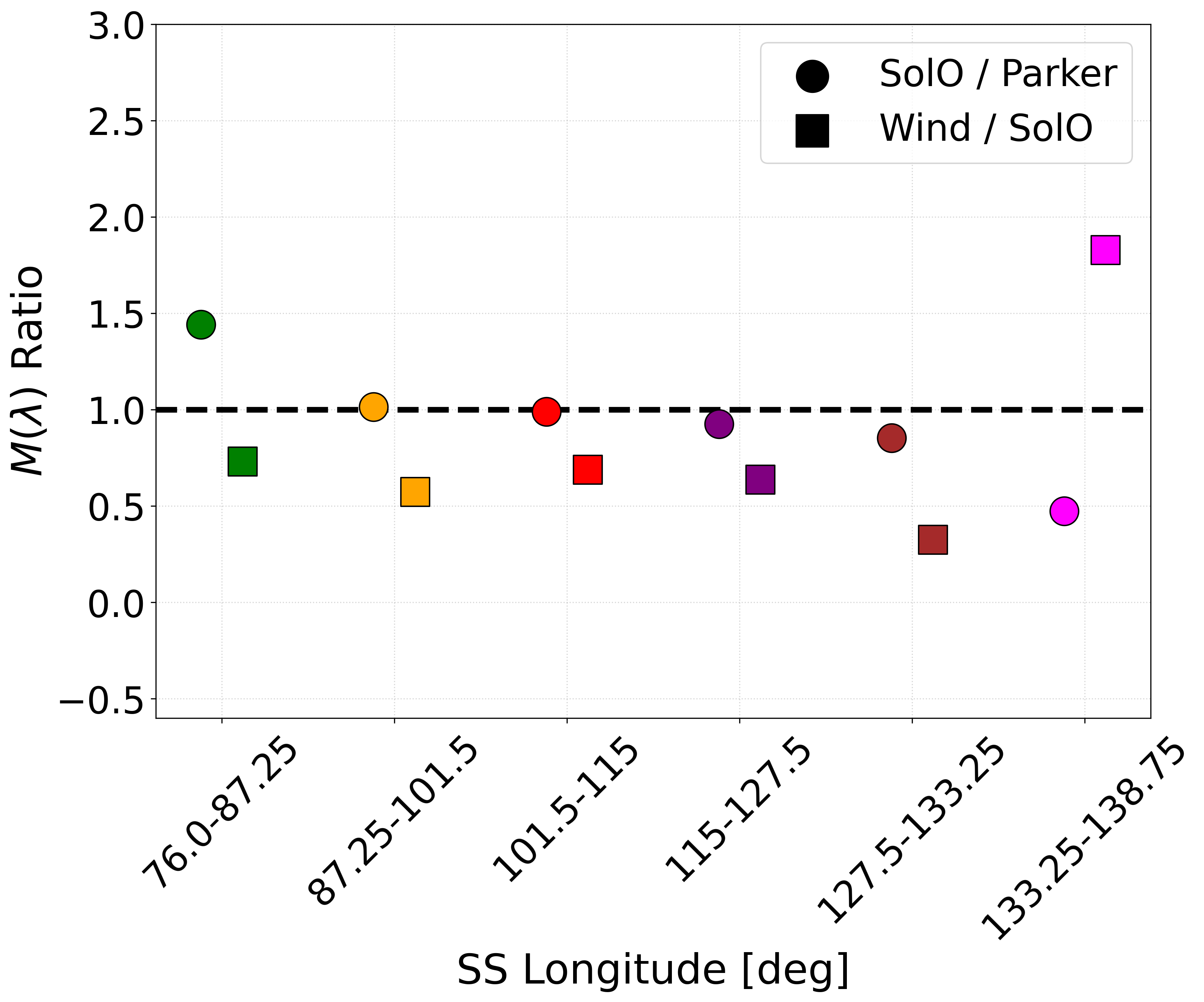}

\vspace{0.1cm}

\includegraphics[width=0.34\textwidth, trim=20 0 20 10, clip]{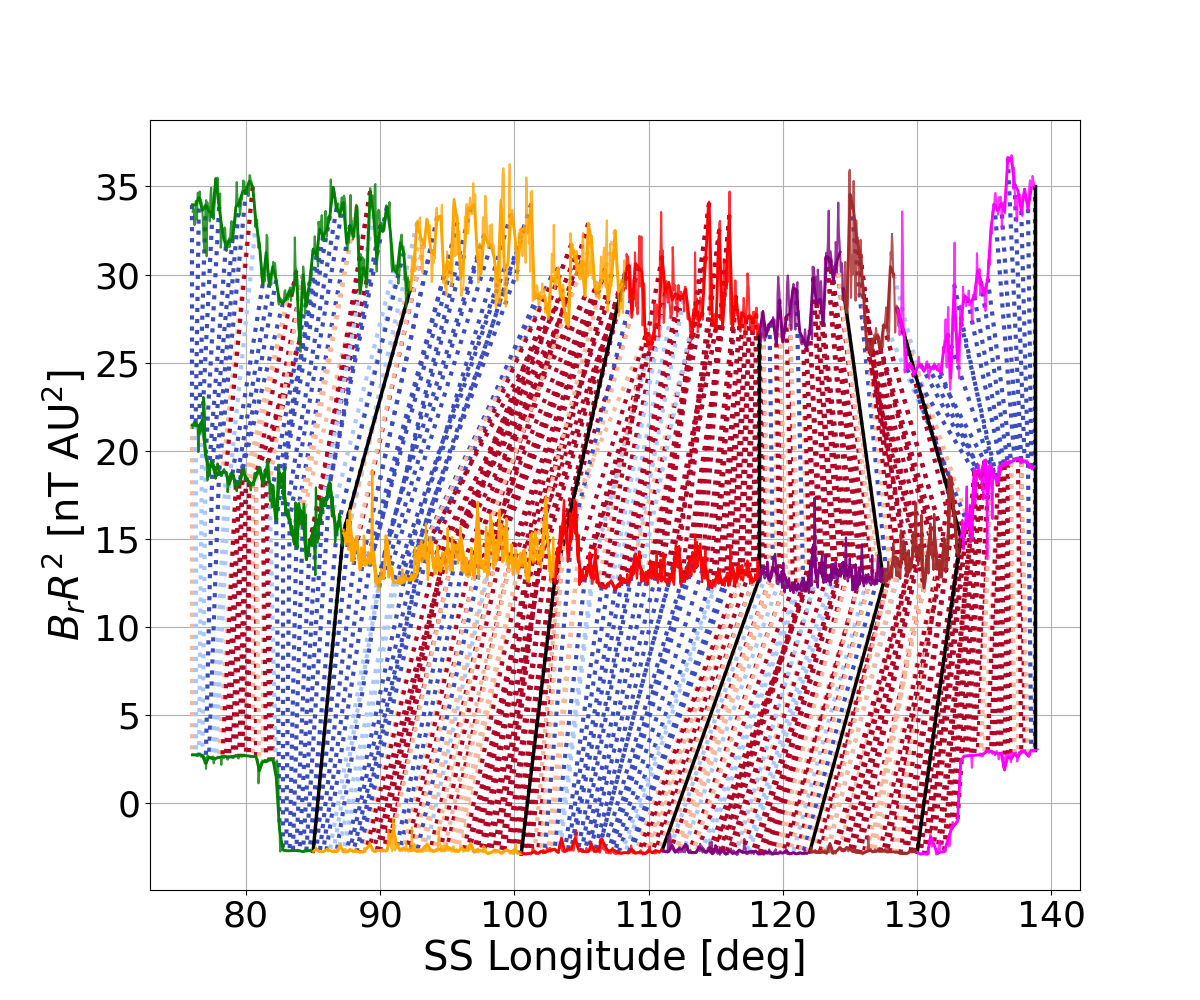}%
\includegraphics[width=0.34\textwidth, trim=20 0 20 10, clip]{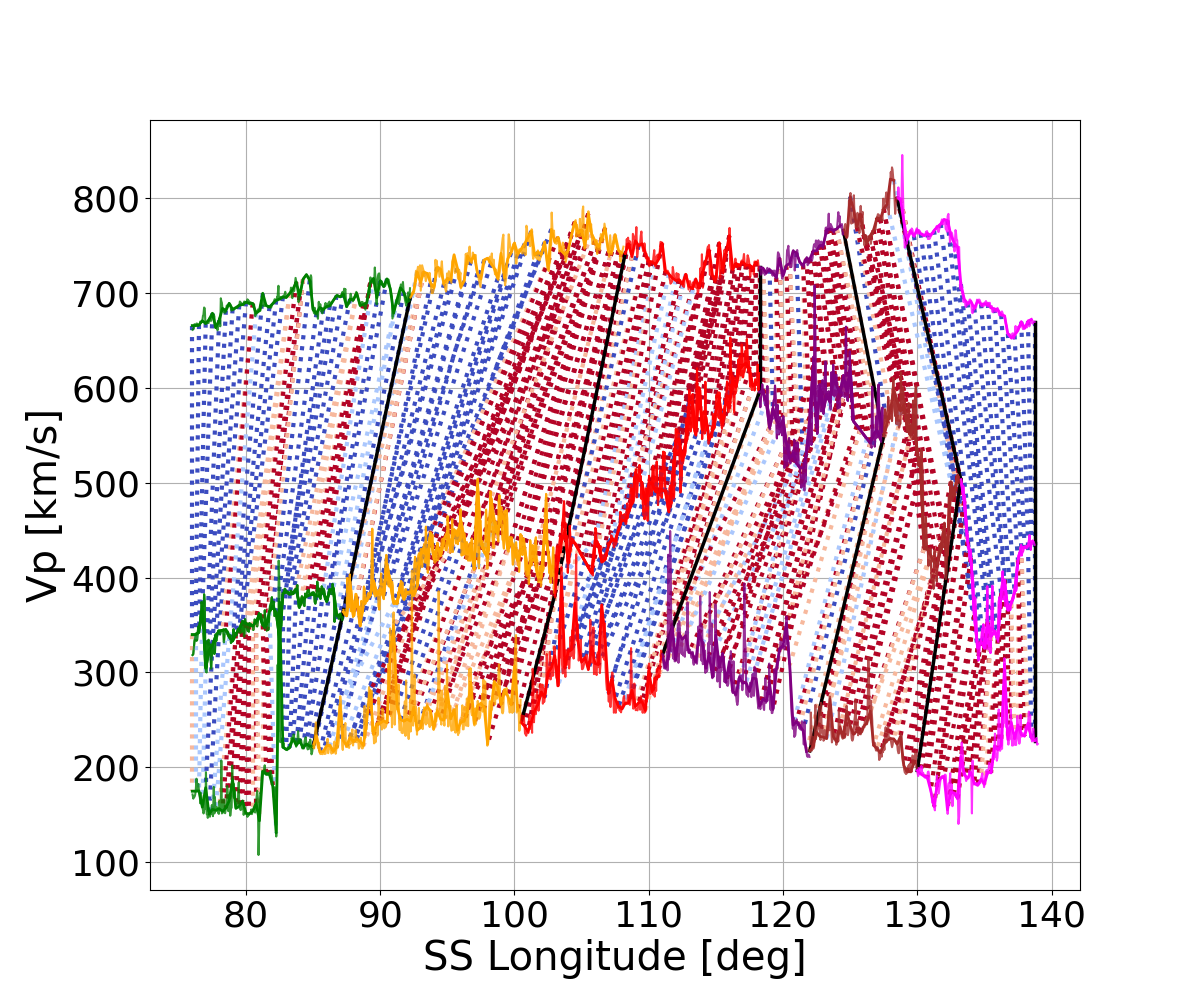}%
\includegraphics[width=0.32\textwidth, trim=0 0 0 0, clip]{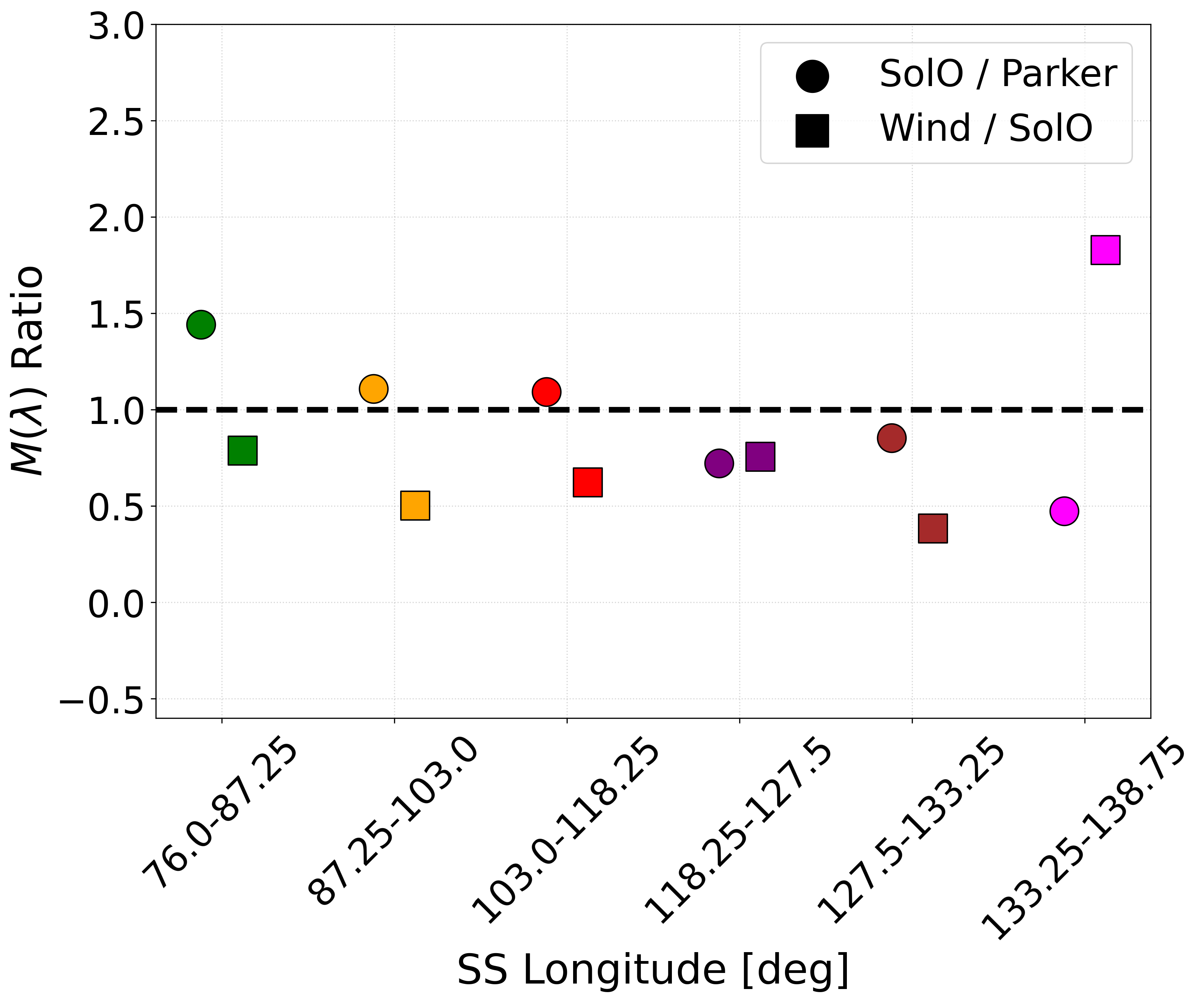}

\caption{Comparison of three methods used to define the boundaries of the mapped SB patches in source surface longitude. From top to bottom: the original visually identified patch identification, boundaries determined from mass flux conservation, and boundaries determined using the optimized Pi-DTW mappings. Pi-DTW mapping lines are in red and blue. Black lines connect the boundaries estimated by each method.
Left column: scaled radial magnetic field, $B_RR^2$ for each method, as functions of source surface longitude. 
Middle column: proton velocity, $V_p$ for each method, as functions of source surface longitude. 
Right column: ratios of the maximum cumulative mass flux between SolO and {\it{Parker}} as well as between \textit{Wind} and SolO for each DTW mapped patch. The horizontal dashed line at one indicates equal cumulative mass flux between the spacecraft.}
\label{Figure 5.}
\end{figure*}

Figure \ref{Figure 4.}b shows the total energy flux as a function of longitude for {\it{Parker}}, SolO and \textit{Wind}. The total energy flux is the sum of all the individual energy components. 

Assessing both $g$ and energy flux, at large scale, the overall stream is approximately conservative at all measurement points. For the left half of the stream as crossed by \textit{Wind}, a lower mass flux (which, as discussed above, was reconcilable by considering the magnetic flux in tandem) manifests in the energy flux trace too, while the right half remains comparable. When energy fluxes are compared on a patch by patch basis later, we normalize them by their mass flux to account for this departure from spherical expansion.

\subsection{Mapping Corresponding Structures with Dynamic Time Warping}

As discussed above, we seek a better method than ballistic mapping to investigate the most plausible mapping from the {\it{Parker}} SB patches to the substructure measured by SolO and \textit{Wind} further out. Initial "by-eye" labeling yielded discrepant and (as will be shown below), highly non-conservative results. 

We therefore implemented a DTW method to map corresponding structures between spacecraft. The DTW is an algorithm that allows for stretching and compression to align similar structures across different time series and is ideal for determining an optimal mapping for the same regions of interest one to one between spacecraft/different distances. Unlike simple ballistic mapping, DTW allows corresponding structures to undergo local expansion or contraction while preserving their overall ordering (i.e. streamlines do not cross). This is particularly important because neighboring solar wind streams interact during radial propagation.

Before applying DTW, the input data were preprocessed using median filtering and normalization. This is important since the DTW method is sensitive to careful filtering, detrending and normalization of the input data. DTW works by mapping a query (i.e. a SolO parameter) onto reference (i.e. {a \it{Parker}} parameter). The cost function for DTW measures the dissimilarity between two points in the reference and query series. More dissimilar points are penalized using higher costs. This drives the DTW algorithm to find mapping configurations which maximize pointwise similarity across time series (in this case longitude series). The algorithm then searches for the warping path that minimizes the cumulative cost across the full interval. The cost function optimization is performed using recursion formulae that are unique to the chosen \textit{step-pattern} which is what constrains the transitions allowed during the optimal path finding process.  

Applying DTW to individual patches separately resulted in overfitting, whereas mapping the entire current sheet bounded interval provided the strongest results.

DTW was also applied on a longitude series of $V_p$ and $B_RR^{2}$ as well as total energy flux. The patches in the total energy flux plots have a greater visual similarity than those in the individual $V_p$ and $B_RR^{2}$ plots, making them easier for DTW to align and identify corresponding structures. Therefore, the total energy flux longitude series were selected as the primary DTW input. An example plot of the DTW mapping applied to total energy flux is shown in Appendix Figure \ref{fig:appendix_dtw}. 

\subsubsection{Physics Informed Dynamic Time Warping (Pi-DTW)}
After creating the DTW mappings by mapping total energy flux in isolation, DTW was further optimized by adding additional physical parameters to the DTW cost function to make it Physics Informed. The cost function works by combining similarity between longitude series with consistency in $B_R$ polarity, |B|, mass flux, energy flux, and $V_p$. The input series were normalized, and the contributions were assigned equal weights. Pi-DTW finds the warping path that best optimizes all quantities at the same time. The optimal Pi-DTW path is therefore the path that minimizes the combined similarity and physics based penalty. Energy flux and mass flux in particular are critical since they are a metric for conservative behavior in the final mapping. 

Although developed here for an investigation of the evolution of switchback patches, the overall methodology is independent of the solar wind structure. In principle, any conserved quantity may be used to align structures across multiple spacecraft.

\subsubsection{Evaluation of the Mapping}

To evaluate the physical consistency of the mapping, we compared three independent estimates of the patch boundaries: (1) the original visually identified boundaries, (2) boundaries refined using mass flux conservation, and (3) boundaries obtained from DTW/Pi-DTW. The comparison is shown in Figure \ref{Figure 5.}. In the mass flux based patch refinement method, the cumulative mass flux was calculated as a function of source surface longitude within each patch, and the longitudinal boundaries were manually adjusted to bring the total mass fluxes of corresponding patches into closer agreement between {\it{Parker}}, SolO, and \textit{Wind}. 

Figure~\ref{Figure 5.} shows the progression of the patch identification from the visually identified boundaries in the top row, to the manually refined mass flux conservation boundaries in the middle row, and finally to the optimized Pi-DTW boundaries in the bottom row. The first column is the scaled magnetic field, second column is the speed, and last column shows the cumulative mass flux ratio for each SB patch between SolO to {\it{Parker}} and \textit{Wind} to SolO, colored in the previous two columns of that row. In each of the mapped quantities of the figure, the bottom series is {\it{Parker}} whose patch shading is fixed, the middle series is SolO, and top series is \emph{Wind} which have been manually shifted in the y-axis for visibility. The blue/red connecting paths show the Pi-DTW mapping between {\it{Parker}} and SolO while the black lines are the coloring paths from each coloring method. The red Pi-DTW lines indicate angular compression going from {\it{Parker}} to SolO to \textit{Wind} while blue indicates rarefaction required to align the measurements. The {\it{Parker}} to SolO plot was plotted separately from the SolO to \textit{Wind} plot, and they were stacked on top of each other. The third column evaluates the ratio of the mass flux integral from each individual mapped patch, and a horizontal black line indicates what is expected for conservative behavior.

In the top row, the visually identified boundaries provide an initial estimate of the locations of the four patches but are strongly discrepant with the final Pi-DTW warping path and have very non-conservative ratios. In the second row, refining these boundaries using mass flux conservation improves the agreement of the corresponding patch mass fluxes between spacecraft. The final Pi-DTW method produces patch boundaries that are close to those obtained from the manual mass flux conservation method, and correspondingly does not noticeably worsen (or improve) the mass flux ratios. Therefore achieving physically defensible results while providing a robust Automated method of estimating patch boundaries without requiring manual intervention for patch identification for each spacecraft. 

Incorporating physical constraints to optimize DTW was important because without it produced mappings that were inconsistent with the conservation properties of the plasma. Pi-DTW preferentially aligned regions with consistent magnetic polarity, similar mass flux, comparable energy flux, and similar proton speed behavior because these were terms added to the cost function. The resulting correspondence therefore combines profile similarity with physically conserved plasma properties, producing a more robust identification of corresponding structures across spacecraft.

The DTW/Pi-DTW mapping derived from the total energy flux profiles was applied to the $V_p$ and $B_RR^2$ plots in order to evaluate how the mapped boundaries correspond to the original plasma and magnetic field properties.

\subsection{Tracking the Radial Evolution of Individual Patches}

Using the final Pi-DTW mappings presented in the bottom row of Figure \ref{Figure 5.}, the plasma and magnetic properties of each mapped SB patch are tracked from {\it{Parker}} to SolO to \textit{Wind} as a function of heliocentric distance. The quantities investigated include each of the parameters listed in Appendix Table \ref{tab:appendix_table}, and plotted in Figure \ref{Figure 6.}, which are helium abundance, plasma $\beta$, cross helicity ($\sigma_C$), switchback associated magnetic fluctuation power, $\delta B^2$, proton speed, velocity enhancement from background, proton temperature, alpha-proton differential streaming, mass flux, energy flux, and longitudinal extent. The observed radial evolution is then compared with previous observational studies of microstreams \citep[e.g.,][]{Neugebauer1995} and theoretical expectations for solar wind expansion and evolution.
The selected quantities were chosen to test specific observational properties of SB patches and microstreams discussed in the introduction section.

Normalized cross helicity and switchback associated magnetic fluctuations quantify the Alfv\'enicity and Alfv\'en wave amplitude of the SB patches as they evolve. Proton speed and the velocity enhancement relative to the local background speed show the corresponding acceleration/deceleration of each SB patch. Proton temperature and plasma beta characterize their thermal evolution, while alpha-proton differential streaming and helium abundance provide additional diagnostics of the plasma state and source.
Mass flux and total energy flux track the conservation of plasma content and energy, and the longitudinal extent quantifies patch dependent compression and rarefaction. We note that mass flux normalization (independently for each patch) is applied to the reported energy flux numbers, following \citet{Rivera2024}.

The patch averaged plasma and magnetic quantities were calculated within the final Pi-DTW defined longitudinal boundaries for each mapped structure. Proton speed, density, temperature, and magnetic field magnitude were obtained directly from the spacecraft measurements.

The helium abundance was defined as,

\begin{equation}
A_{\rm He}
=
\frac{n_{\alpha}}{n_p}\times100\%,
\end{equation}

where $n_{\alpha}$ and $n_p$ are the alpha-particle and proton number densities, respectively. 

The proton plasma beta was calculated from the ratio of the proton thermal pressure to the magnetic pressure,

\begin{equation}
\beta
=
\frac{n_pk_BT_p}
{B^2/(2\mu_0)}.
\end{equation}

Where each pressure term is in units of $\mathrm{J/m^{3}}$. The resulting plasma beta values were averaged within each mapped patch.

The normalized cross helicity, $\sigma_c$, was calculated to characterize the Alfv\'enicity of the structures. It was calculated as,

\begin{equation}
\sigma_c
=
\frac{
2\langle\delta\mathbf{v}\cdot\delta\mathbf{b}\rangle
}{
\langle|\delta\mathbf{v}|^2\rangle+
\langle|\delta\mathbf{b}|^2\rangle
}.
\end{equation}
where $\delta{v}$ and $\delta{b}$ represent the velocity and magnetic field fluctuations, respectively. The fluctuations were calculated for each RTN component by subtracting the corresponding rolling mean, with a 10 minute widow, from the measured values. 

The switchback associated magnetic fluctuation power series was calculated from the magnetic field fluctuations as $\delta{B}^2$, and subsequently averaged within each mapped patch.

To quantify the velocity enhancements, the characteristic `arc' shape of microstreams and switchback patches \citep[e.g.][]{Neugebauer1995}, $V_p$ measurements were smoothed using an 11 point rolling median, with a minimum of three valid measurements per window. Within each mapped patch, the maximum and minimum of this rolling median profile were identified as

\begin{equation}
\Delta V
=
V_{p,\rm max}
-
V_{p,\rm min}.
\end{equation}

The alpha-proton differential streaming was calculated from the
difference between the alpha and proton velocities as:

\begin{equation}
\Delta V_{\alpha p}
=
V_{\alpha}-V_p,
\end{equation}

and was averaged within each mapped patch where alpha-particle
measurements were available i.e {\it{Parker}} and \textit{Wind}.

The mass flux was within each SB patch was computed as discussed in Section \ref{subsec:mass_energy_constraints}.

The kinetic, enthalpy, gravitational, and Alfv\'en wave energy flux components were calculated using Equation~\ref{eq:energy_flux}, and the total energy flux was defined as,

\begin{equation}
W_{\rm T}
=
W_\text{Kinetic}+W_\text{Enthalpy}+W_\text{Alfvén~Wave}+W_\text{Gravitational}.
\end{equation}

The individual energy components and total energy flux were averaged within each patch. For the radial profile comparison, the SolO and \textit{Wind} energy components were normalized by the mass flux.

Finally, the longitudinal extent of each structure was calculated from

\begin{equation}
\Delta\lambda
=
\lambda_{\rm end}
-
\lambda_{\rm start},
\end{equation}

using the final Pi-DTW defined source surface boundaries. Changes in $\Delta\lambda$ between spacecraft provide an empirical measure of the longitudinal compression or rarefaction of each mapped structure.

Radial profiles of the most important of these parameters is also shown in Figure \ref{Figure 7.} and are contextualized with the empirically expected radial profiles using an ``iso-poly + forcing'' Parker solar wind model. These profiles are generated using the ``iso-poly'' framework \citep{Dakeyo2022} with an additional momentum from Alfv\'enic fluctuation energy, described as the ``iso-poly + forcing'' model \citep{Shi2022, Rivera2024, Rivera2025, Badman2025}, implemented using the \texttt{ParkerSolarWind} \citep{ParkerSolarWind} package.

\section{Results and Discussion}
\label{sec:Results and Discussions}

We track the four main patches identified in the $B_RR^{2}$ and $V_p$ series at {\it{Parker}}, shown in Figure \ref{Figure 3.}, to SolO and \textit{Wind}, colored the same way. 

Having established that the total mass flux normalized by the magnetic flux is approximately conserved within the region bounded by the two heliospheric current sheets (Figure~\ref{Figure 4.}a), and that the total energy flux is similar across the stream from spacecraft to spacecraft (Figure \ref{Figure 4.}b), substructure in the total energy flux longitude series is observed at the patch scale with qualitatively similar features in certain spacecraft pairs. The very presence of coherent substructure suggests that conservation of energy flux on the patch scale can be incorporated in order to improve the evolutionary DTW mapping and motivates the use of this quantity as the primary mapping key, rather than $B_R$ and $V_p$ the archetypal SB patch/microstream quantities, which themselves evolve strongly and do not intrinsically enforce any physical conservation.

\subsection{Physical Insight from Optimized Mapping Between Patches}

The final multi-spacecraft DTW mapping was presented in Figure \ref{Figure 5.}, showing the results of the optimized Pi-DTW mappings. Besides functioning as the basis for connecting and studying the evolution of patches, the mapping results offer some physical insight in their own right: By colorizing the warping path according to compressions (red) and rarefactions (blue), we observe rarefaction on the left side CS crossing (trailing edge) and a compression on the right side CS crossing (leading edge) consistent with Co-rotating Interaction Regions (CIRs) (i.e. an asymmetry consistent with the expected influence of the Parker spiral). 

Similar behavior is present within several of the individual SB associated substructures, most prominently the orange patch, where compression and rarefaction occur on opposite sides of the structure. However, because the spacecraft sample different heliographic latitudes, particularly on the right hand side of the interval, part of the apparent longitudinal evolution may also reflect latitudinal structure rather than purely radial evolution.

\subsection{Patch Investigation}
\begin{figure*}[ht!]
\centering
\includegraphics[width=0.7\textwidth]{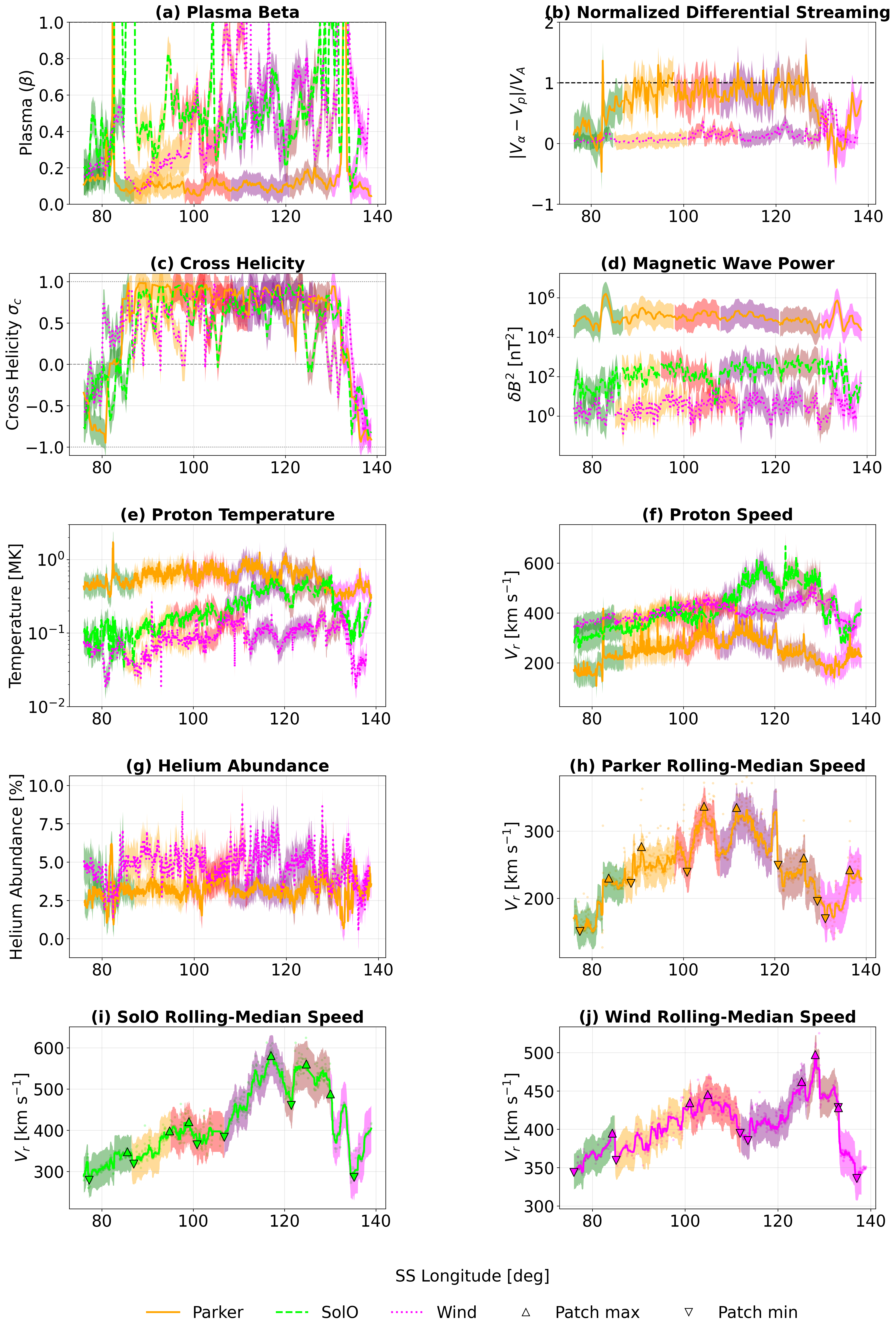}
\caption{Multi-spacecraft comparison of SB patch plasma and magnetic field properties. Plasma and field properties are shown as a function of source surface longitude for each mapped patches observed by {\it{Parker}}, SolO, and \textit{Wind}. Panels show (a) plasma beta, (b) proton–alpha differential streaming normalized by the Alfv\'en speed, (c) cross helicity, (d) magnetic fluctuation power, (e) proton temperature, (f) proton speed, (g) helium abundance, and (h–j) rolling median max-min proton velocity for {\it{Parker}}, SolO, and \textit{Wind} used to calculate $\Delta V$, respectively. Shaded colors identify the corresponding patches, while spacecraft are distinguished by line color (orange for {\it{Parker}}, lime for SolO, and magenta for \textit{Wind}) and line style (solid for {\it{Parker}}, dashed for SolO, and dotted for \textit{Wind}). Upward and downward pointing triangles indicate the maximum and minimum velocities within each patch, respectively.}
\label{Figure 6.}
\end{figure*}

Using the correspondences between the plasma structures observed at {\it{Parker}}, SolO, and \textit{Wind} obtained from Pi-DTW, we next examine the radial evolution of the plasma and magnetic properties within each mapped patch. These properties are summarized in Figures \ref{Figure 6.}, \ref{Figure 7.}, and Appendix Table \ref{tab:appendix_table} which includes a complete list of the investigated parameters and their patch-averaged values for each of the patches at each spacecraft. 

\subsubsection{Microstream Characteristics at SolO and \textit{Wind}}
To investigate whether the mapped structures observed farther from the Sun exhibit microstream characteristics, we first compare their properties at SolO and \textit{Wind} with the observational characteristics reported in the literature. The rightmost two columns in Table~\ref{tab:Table2} summarizes these properties and compares them with the observations in this study.

The mapped structures exhibit identifiable mesoscale velocity enhancements at both SolO and \textit{Wind}. Using the rolling median max-min definition for $\Delta V$, the patch amplitudes at SolO range from approximately $56$ to $198~\mathrm{km\,s^{-1}}$, and at \textit{Wind} from approximately $51$ to $77~\mathrm{km\,s^{-1}}$.
Classic microstreams were defined by \citet{Neugebauer1995} as velocity amplitudes of approximately $40~\mathrm{km\,s^{-1}}$ or greater from a running background. Therefore, the persistence of velocity structures at \textit{Wind} is microstream-like. 

The \textit{Wind} patch averages for helium abundance range from approximately $4.3\%$ to $5.3\%$. These values are consistent with typical helium abundances observed in fast or Alfv\'enic slow wind \citep{Alterman2019, Johnson2024} and with the values reported for Ulysses microstreams, where the alpha-to-proton ratio was $0.047\pm0.001$ at velocity peaks and $0.040\pm0.001$ at dips. This is empirically consistent with this being a coronal hole wind \citep{Neugebauer1995}.

We also see a general pattern of compression on one side and a rarefaction on the other side for the individual patches. This pattern provides an additional connection between these mapped structures and classical microstreams. \citet{Neugebauer1995} reported that microstreams are characterized by compression on their leading edges and rarefaction on their trailing edges.
The respective widths of the patches remain on mesoscale angular scales, generally of order $\sim5-10$\textdegree, and they show both expansion and compression depending on the individual patch. Therefore, the mapped structures retain comparable mesoscale longitudinal extent/widths despite patch dependent stretching and compression.

\subsubsection{Radial Evolution of SB Patches into Microstream-like Structures}

Under the hypothesis that near-Sun SB patches evolve into microstreams with increasing heliocentric distance, the mapped structures should progressively acquire the observational properties used to identify microstreams farther from the Sun while remaining traceable as coherent mesoscale structures. Table~\ref{tab:Table2} summarizes the radial behavior implied by this hypothesis based on previous observational studies and compares it with the evolution measured here. 

Several common radial trends are evident across the four mapped patches. Overall, these trends are broadly consistent with the evolution of the SB patches' observational properties into that of microstreams reported in the literature, such as in \citet{Neugebauer1995, Horbury2023, Soni2024, Kumar2023, Huang2023}.

For example, the structures exhibit increasing bulk flow speeds, decreasing Alfv\'enicity, reduced alpha-proton differential streaming, and the persistence of velocity enhancements as they propagate from {\it{Parker}} to SolO and ultimately to \textit{Wind}. The longitudinal extents of the patches remain roughly consistent while they undergo patch dependent compression and rarefaction on smaller scales. 

The mean proton speed increases from {\it{Parker}} to SolO in all four patches. From SolO to \textit{Wind}, the evolution is mixed: orange and red continue to increase to $393$ and $428~\mathrm{km\,s^{-1}}$, whereas purple and brown exhibit lower patch averaged speeds at \textit{Wind} than at SolO but remain substantially faster than they were at {\it{Parker}}, at $419$ and $454~\mathrm{km\,s^{-1}}$, respectively. 

The cross helicity evolution is not identical for all four structures. The patch averaged normalized cross helicity generally decreases or remains comparable with increasing heliocentric distance. The orange patch decreases from $0.91$ at {\it{Parker}} to $0.68$ at SolO and $0.53$ at \textit{Wind}, the red patch changes from $0.86$ to $0.68$ to $0.66$, and the brown patch decreases from $0.72$ to $0.61$ to $0.27$. The purple patch differs from this pattern, with values of $0.80$, $0.72$, and $0.81$, respectively. The significance of these differences is evaluated relative to the within patch variability reported in Table~\ref{tab:appendix_table} from $0.09$ to $0.52$, indicating substantial variability within the structures. In particular, the apparent increase in cross helicity for the purple patch from $0.72$ at SolO to $0.81$ at \textit{Wind} is small compared to the within patch variability and therefore does not provide compelling evidence for a systematic increase in Alfv\'enicity with heliocentric distance.

\begin{figure*}[ht!]
\centering
\includegraphics[width=0.7\textwidth]{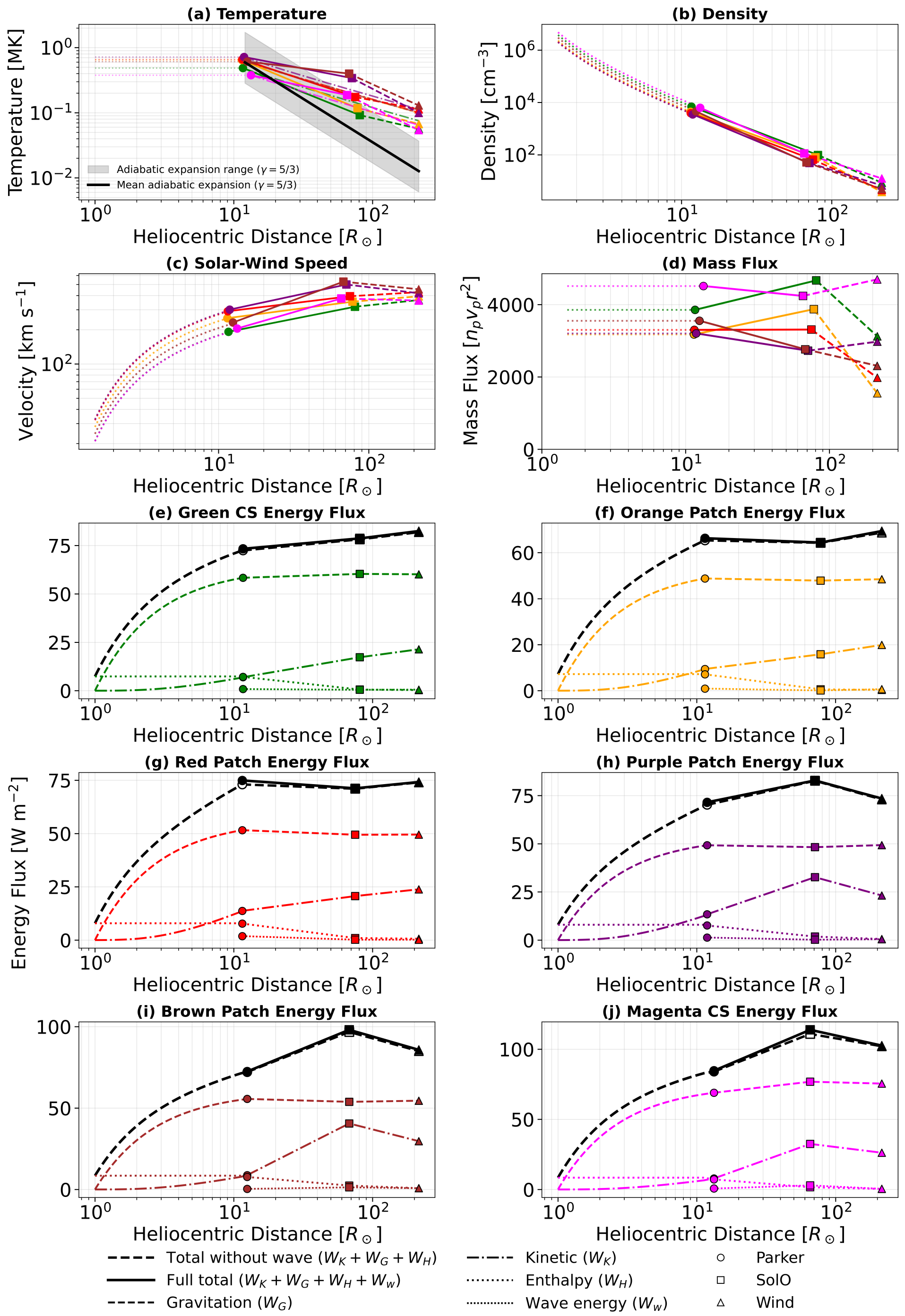}
\caption{Panels (a–d) show the radial evolution of proton temperature, density, proton speed, and mass flux for the two CS crossings and the four mapped patches. Panels (e–j) show the corresponding energy flux plots for each patch and CS crossing, decomposed into kinetic, gravitational, enthalpy, and wave contributions, together with the total energy flux with and without the wave term. Marker shape indicates spacecraft location: {\it{Parker}} (circles), SolO (squares), and \textit{Wind} (triangles).}
\label{Figure 7.}
\end{figure*}

The switchback associated magnetic fluctuation power ($\delta B^2$) decreases dramatically with heliocentric distance. Correspondingly, the total wave energy flux exhibits a more modest reduction as can be seen in Figure \ref{Figure 7.}.

The velocity enhancement ($\Delta V$ from background) also remains identifiable after the wave activity has weakened. Using the rolling median max-min definition used to compute the values for $\Delta V$, the patch amplitudes at {\it{Parker}} range from approximately $55$ to $98~\mathrm{km\,s^{-1}}$. As discussed in the previous subsection, mesoscale velocity variations remain identifiable at SolO and \textit{Wind}, with amplitudes ranging from $56$ to $198~\mathrm{km\,s^{-1}}$ and $51$ to $77~\mathrm{km\,s^{-1}}$, respectively. The most important result is therefore not a monotonic increase or decrease in $\Delta V$, but the persistence of a mesoscale velocity variation at \textit{Wind} after the switchback associated magnetic fluctuation power has decreased by several orders of magnitude. 

The reduction in Alfv\'enic wave activity occurs while the velocity enhancement of the structures survives.
This combination is important for the proposed SB Patch to microstream evolution. The magnetic wave activity becomes progressively weaker, while the velocity enhancement survives. \citet{Soni2024} similarly found that microstreams observed farther from the Sun contained fewer switchbacks than near-Sun SB Patches and that the proton speed was approximately $10\%$ greater than the surrounding pristine wind, which they interpreted as being consistent with conversion of switchback magnetic energy into kinetic energy of the plasma. However, we note that for the patches studied here, the wave energy flux is a relatively small component of the total energy flux and therefore, while this is an important characteristic of this evolution, here it is unlikely to be dynamically important to the patch to microstream evolution \citep[see also][]{Rivera2026_eclipse}. 

The patch averaged alpha-proton differential streaming decreases substantially with heliocentric distance, from approximately $98$-$114~\mathrm{km\,s^{-1}}$ at {\it{Parker}} to $4$-$15~\mathrm{km\,s^{-1}}$ at \textit{Wind}. The near-Sun values are comparable in magnitude to previously reported differential streaming values of $150 \pm 50~\mathrm{km\,s^{-1}}$ in Alfv\'enic wind close to the Sun in \citet{Wang2025}.  

All four {\it{Parker}} patches have low plasma beta values of approximately $0.09$-$0.13$, indicating strongly magnetically dominated plasma. At SolO, the patch averages increase to approximately $0.49$-$0.59$, and at \textit{Wind} they range from $0.23$ to $0.75$. Given the intra-patch standard deviations of $0.05$-$0.17$ at SolO and $0.05$-$0.27$ at \textit{Wind}, the individual differences between the SolO and \textit{Wind} values should be interpreted cautiously. It is generally evident from the longitude profiles in Figure \ref{Figure 6.} that there is a transition from low $\beta$ plasma near {\it{Parker}} to generally higher and more variable $\beta$ farther from the Sun, consistent with their transition from highly Alfv\'enic and non-compressive close to the Sun, to predominantly hydrodynamic structures further out. This radial increase is consistent with the general evolution of the solar wind.

The {\it{Parker}} patch averages for helium abundance are between approximately $3.1\%$ and $3.5\%$, while the \textit{Wind} values range from approximately $4.3\%$ to $5.3\%$. As mentioned in the \ref{sec:Mapping Methodology} section, SolO helium abundance is not reported because a measurement was unavailable, and therefore, where required for mass and energy flux calculations, a helium abundance of $4\%$ was assumed. 

Given that there are likely to be higher measurement uncertainties at {\it{Parker}} due to partial field of view effects, we do not interpret this difference as evidence for radial evolution of the helium abundance. A more sophisticated treatment of alpha and proton velocity distribution functions in future work is warranted.

Figure \ref{Figure 7.} plots the patch averaged measurements in a radial context and compares the observations with the expected radial profiles from the statistically constrained iso-polytropic/ParkerSolarWind model described in the methods section.
The density profiles broadly follow the strong radial decline expected from spherical expansion, while the velocity profiles show that all four structures undergo substantial acceleration between {\it{Parker}} and SolO. From SolO to \textit{Wind}, the orange and red patches show further increases in the mapped patch averaged speed, whereas purple and brown show an implied decrease in speed. Because the latter structures are sampled across larger latitudinal separations, these differences may reflect latitudinal or temporal structure in addition to radial evolution.

\startlongtable
\begin{deluxetable*}{l l l l l}

\tablecaption{
\label{tab:Table2}
Comparison of the observed radial evolution with expected microstream
properties from previous studies.
}
\tablehead{
\colhead{Parameter} &
\colhead{\shortstack{Expected\\Evolution}} &
\colhead{\shortstack{Observed\\Evolution}} &
\colhead{\shortstack{Expected\\in MS}} &
\colhead{\shortstack{Does Our Observation\\Match the Expected?}}
}
\startdata
He Abundance (\%) &
Conserved &
\shortstack[l]{Growing \\--measurement limitations} &
\shortstack[l]{4.7\% peaks\\4.0\% dips} &
Yes\\
\tableline
Plasma $\beta$ &
Increase &
Increase &
\shortstack[l]{ enhanced} &
\shortstack[l]{Yes}\\
\tableline
Cross Helicity &
\shortstack[l]{Weak\\decrease} &
\shortstack[l]{Weak\\decrease} &
\shortstack[l]{Lower\\Alfv\'enicity} &
Yes, generally\\
\tableline
Magnetic Fluctuation Power &
Decrease &
Decrease &
\shortstack[l]{Reduced\\$\delta B^2$} &
Yes\\
\tableline
Velocity (km s$^{-1}$) &
\shortstack[l]{Increase\\Parker--SolO\\Flat\\SolO--\textit{Wind}} &
\shortstack[l]{Increase\\Parker--SolO\\Mixed\\SolO--\textit{Wind}} &
\shortstack[l]{Persistent\\velocity\\enhancement} &
Yes\\
\tableline
$\Delta V$ (km s$^{-1}$) &
\shortstack[l]{Increase\\or\\decrease} &
\shortstack[l]{Patch\\dependent} &
\shortstack[l]{$\sim20$ km\\s$^{-1}$\\or larger} &
Yes\\
\tableline
Differential Streaming (km s$^{-1}$) &
\shortstack[l]{Decrease\\with\\distance\\with $V_A$} &
\shortstack[l]{Decrease} &
\shortstack[l]{$150\pm50$ km s$^{-1}$\\close to \\the Sun, Small\\farther\\from Sun} &
Yes\\
\tableline
Longitudinal Extent \textdegree &
\shortstack[l]{Approximately\\similar} &
\shortstack[l]{Approximately\\similar} &
\shortstack[l]{$\sim5$--$10$\textdegree} &
Yes\\
\tableline
Mass Flux ($\mathrm{cm^{-3}\,km\,s^{-1}\,AU^2}$) &
\shortstack[l]{Approximately\\conserved} &
\shortstack[l]{Approximately\\conserved} &
--- &
---\\
\tableline
Energy Flux (W m$^{-2}$) &
\shortstack[l]{Polytropic\\evolution} &
\shortstack[l]{Polytropic\\evolution} &
--- &
---\\
\tableline
Temperature (MK) &
\shortstack[l]{Decrease} &
Decrease &
\shortstack[l]{Velocity peak\\$\sim6\%$ faster\\$\sim12\%$ hotter} &
Yes\\
\enddata
\end{deluxetable*}

The temperature profiles in Figure \ref{Figure 7.}a show that all patches cool strongly with heliocentric distance, from approximately $0.61$-$0.72~\mathrm{MK}$ at {\it{Parker}} to $0.12$-$0.40~\mathrm{MK}$ at SolO and $0.07$-$0.13~\mathrm{MK}$ at \textit{Wind}. However, comparison of the observed radial profiles (solid lines) with the expected radial profiles (dotted lines) again from the iso-polytropic solution, shows that several structures remain hotter than expected from adiabatic expansion (black solid line), implying that, as with the typical solar wind behavior, the patches receive some additional heat input to counter the cooling from expansion during their propagation.  The gray shaded region around the solid black line (mean adiabatic expansion) represents the range of expected adiabatic cooling ($\gamma=5/3$). 
The original Ulysses microstream study found that velocity peaks were approximately $6\%$ faster and $12\%$ hotter than adjacent dips, establishing an association between enhanced flow speed and enhanced proton temperature. We similarly observe temperature enhancements associated with microstream structures, with the corresponding temperature differences, $\Delta T = T_{\max}-T_{\min}$, reported in Table \ref{tab:appendix_table}.

The total energy flux, which is marked by the black lines in \ref{Figure 7.}, is more consistent across the mapped structures than any individual component, which provides physical support for using total energy flux as the primary quantity in the DTW mapping. The total flux is also barely altered by the inclusion of the wave energy flux, indicating it is a minority component in this stream. 

\section{Summary and Conclusions}
\label{sec:Summary and Conclusions}

In summary, several observational features are worth noting. First, we have verified that the properties of solar wind patches identified visually close to the Sun (at {\it{Parker}}) and connected to observations further out using our Pi-DTW method do indeed evolve to exhibit the general properties reported in the Ulysses microstream observations in \citet{Neugebauer1995}. Second, inspection of the velocity, magnetic field, and total energy flux profiles across multiple spacecraft traversals reveal the same number of sub-structures/patches at each spacecraft. Third, the DTW and Pi-DTW mappings preserve the identity of the patch, producing a connection that is consistent with conservation laws and whose radial evolution can be subsequently studied. These findings support the hypothesis that mesoscale structures can survive and preserve a traceable mapping to their inner heliospheric counterparts under some circumstances.

Overall, the left hand side of the interval appears to provide the most reliable mapping between spacecraft. This is evident from the asymmetry observed in the velocity profiles, which is significantly larger on the right hand side of the interval than on the left, making the left hand structures better constrained for spacecraft-to-spacecraft comparison. Additionally, the radial energy flux profiles in Figure \ref{Figure 7.} show the same asymmetry, where the left side patches (orange and red) match the expected profiles and follow the radial profiles in \citet{Rivera2024} while the right side patches (purple and brown) deviate slightly, mainly in their kinetic energy terms at Solar Orbiter which are notably larger and out of family. 

One alternative explanation we considered for the stronger acceleration between {\it{Parker}} and SolO on the right hand side could be that it resulted from enhanced Alfv\'enic wave energy flux. However, the wave energy flux analysis shown in Figure \ref{Figure 7.} does not show a corresponding enhancement in wave energy flux. Therefore, the observed acceleration cannot be explained solely by the wave energy flux measured here.

\begin{figure*}[ht!]
\centering

\includegraphics[width=0.55\textwidth]{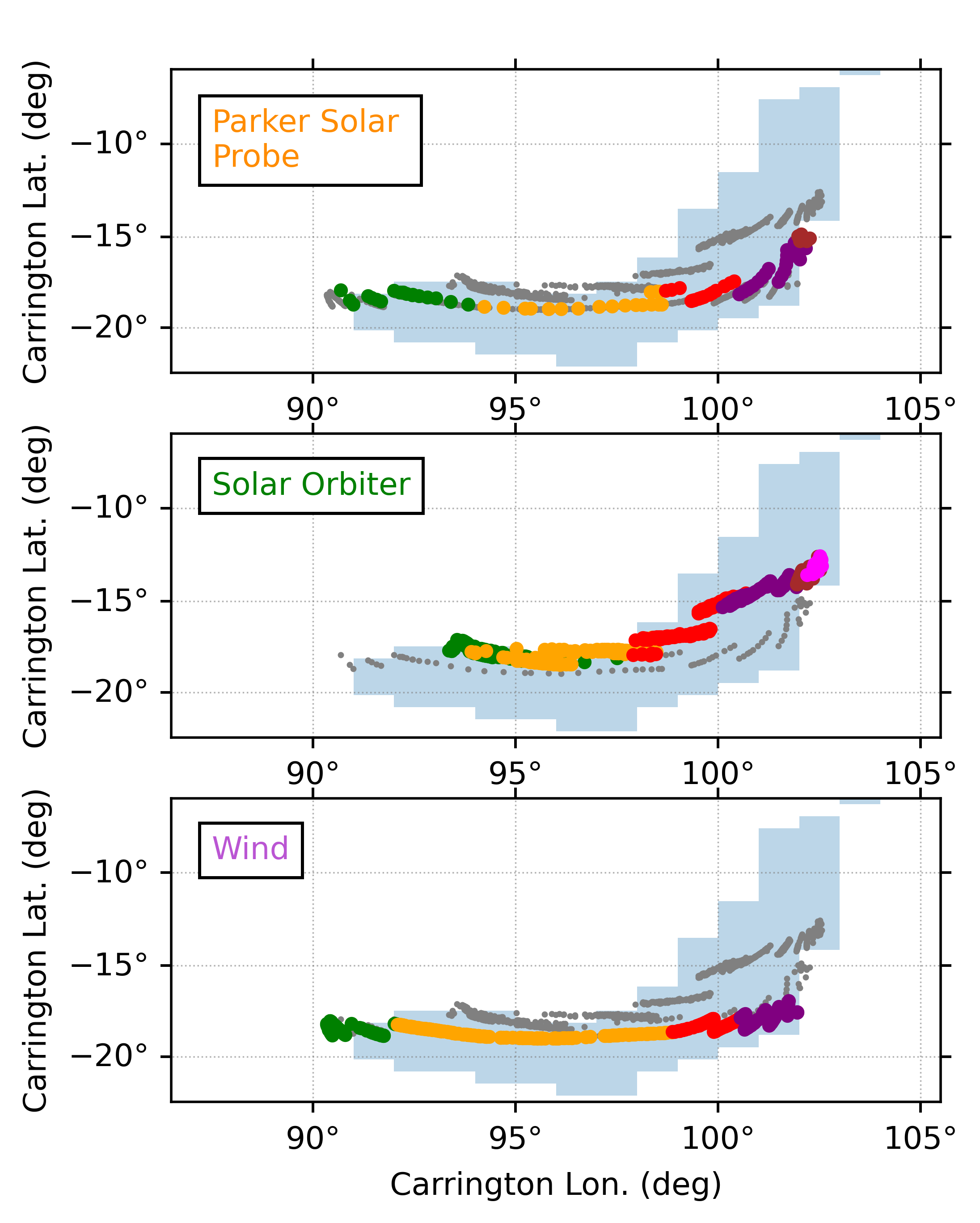}

\caption{A zoom in of the footpoints color coded as before. The {\it{Parker}} and \textit{Wind} footpoints are observed to be essentially indistinguishable, while the SolO ones are displaced slightly upwards and more significantly to the right (at higher Carrington longitudes) corresponding to the purple and brown patches studied in this work.
}
\label{Figure 8.}
\end{figure*}

A more plausible explanation is provided by the mapped footpoints in Figure \ref{Figure 8.} which shows the PFSS-derived coronal hole boundary along with the mapped footpoint locations of each spacecraft. The model shown is as used in Figure \ref{Figure 2.} and \citet{Rivera2026_eclipse}. Although both spacecraft data series traverse the same coronal hole longitudinally, the footpoints are distributed widely in latitude on the right side as shown in Figure \ref{Figure 8.}. On the left side, there is a horizontal edge of the CH which is sensitive to longitude but not latitude, while on the right it is a vertical edge which is insensitive to longitude but sensitive to latitude. This might be because the right most SolO structures originate from a different region of the coronal hole. The discrepancies at SolO in the rightmost patches (purple and brown) may be related, therefore, to the latitudinal structure of the coronal hole. Consequently, the orange and red structures on the left hand side provide the most strongly constrained spacecraft-to-spacecraft comparison, whereas the radial interpretation of the right hand structures is less certain.

The energy component profiles provide a test of whether the observed acceleration can be accounted for by conversion of Alfv\'enic wave energy into kinetic energy. If wave energy dissipation were the dominant source of acceleration between spacecraft, a decrease in the wave energy flux component would be expected to accompany the corresponding increase in kinetic energy flux. For the better constrained orange and red patches, however, the measured change in wave energy flux is not significant, as evident in Figure \ref{Figure 7.}(e)-(j). 
For the left hand patches where the total energy flux is relatively flat (orange and red), the wave energy flux alone is insufficient to explain the observed acceleration and that other energy components, such as thermal pressure, likely play a larger role for this Alfv\'enic slow wind. 

Under the persistent flux tube model used for the reference profiles, the total energy flux should remain approximately conserved with heliocentric distance. The measured total energy flux values are broadly consistent with this behavior, especially for the left most patches. 

Although the switchback associated magnetic fluctuation power decreases dramatically from {\it{Parker}} to SolO and ultimately to \textit{Wind}, the corresponding patch scale velocity enhancements remain throughout. This behavior is consistent with the proposed evolution from wave dominated SB patches to microstream velocity structures.

Finally, it should be emphasized that the present interval samples Alfv\'enic slow solar wind. The observed evolution from strongly Alfv\'enic SB patches to microstream like structures may therefore not apply universally to all solar wind regimes. SB patches are a characteristic of Alfv\'enic solar wind and consequently cannot provide progenitors for microstreams, if present, within non-Alfv\'enic slow wind streams. In contrast, fast coronal hole wind is also strongly Alfv\'enic and commonly contains SB patches with more energetic switchbacks \citep{Halekas2023, Badman2026}, making it an important regime in which to test whether the same evolutionary connection occurs, and in particular, whether the switchbacks are more dynamically important in such a case. Applying the present methodology to a fast coronal hole wind interval, such as that studied recently in \citet{Badman2026}, will therefore provide an important test of the generality of the proposed scenario.

Taken together, the observations demonstrate that mesoscale structures identified near {\it{Parker}} can remain traceable through SolO and to \textit{Wind}, although the confidence of the correspondence varies across the stream. The orange and red structures provide the strongest one-to-one radial correspondences, supported by the Pi-DTW mapping and relatively well aligned source region footpoints. The purple and brown structures retain qualitatively corresponding mesoscale signatures but are less confidently associated because the spacecraft footpoints sample a broader range of coronal hole latitudes. We therefore interpret the left hand structures as the strongest evidence that near-Sun SB patches can evolve with distance to mesoscale velocity structures with microstream like properties, while treating the right hand associations more cautiously. 

We conclude with some key takeaways:

\textbf{Can individual mesoscale structures be followed between spacecraft?}

The combination of visual patch identification, mass flux conservation, total energy flux conservation, and Pi-DTW demonstrates that the same patch structures can be followed from {\it{Parker}} through SolO and to \textit{Wind}. The strong agreement across the entire current sheet bounded interval is first considered rather than immediately trying to map the individual patches, indicating that the stream remains coherent while its substructures evolve under compression and rarefaction.

The approximately conserved total mass flux implies that the plasma is primarily redistributed within the interval. Likewise, the similarity of the total energy flux profiles suggests that energy is redistributed among different components rather than being lost from the structure. Together, these observations support the conclusion that the solar wind stream survives propagation over approximately one month and nearly 1 AU and thus the radial evolution of the individual patches can be tracked.

\textbf{Implications of the DTW mapping}

This study also demonstrates the value of dynamic time warping as a heliophysics analysis tool. Rarefaction and compression are expected during solar wind propagation \citep{Milosic2023}, but identifying structures undergoing these transformations has remained difficult in detail, even for physics models. By following conserved quantities like the total energy flux profile rather than individual velocity or magnetic field profiles, DTW successfully identifies physically meaningful mappings while accounting for longitudinal stretching and compression.

An important outcome of this work is that the DTW mapping itself provides empirical estimates of where compression and rarefaction have occurred during radial propagation, rather than simply aligning two time series. Compression/rarefaction, and turbulence effects etc. play an important role in the plasma heating and energization process \citep{8_Ghanbari2023,9_Borovsky2010}, and identifying how they influence the solar wind dynamics as well as how turbulence transitions from large scale structures to particle heating scales is crucial for solar wind modeling and heliospheric studies. 

The optimized Pi-DTW further improves these mappings by requiring consistency in physical parameters such as magnetic polarity, |B|, mass flux, energy flux, and proton speed. The resulting mappings provide a physically robust method for tracking evolving sub-structures between spacecraft without assuming simple ballistic propagation.

Beyond the present application, this methodology provides a general framework for future multi-spacecraft heliophysics studies, and provide valuable observational constraints for space weather forecast models. The Pi-DTW provides a robust method for tracking evolving plasma structures, reducing dependence on simplified ballistic assumptions.

\textbf{Do Switchback Patches evolve into Microstreams?}

The observations presented here are generally consistent with the hypothesis that the mapped SB patches evolve to microstreams as they propagate outward.

Near {\it{Parker}}, the structures are characterized by strong Alfv\'enic characteristics, including high normalized cross helicity, large alpha-proton differential streaming, and strong switchback associated magnetic fluctuation power. With increasing heliocentric distance, these magnetic and wave characteristics weaken, whereas the velocity enhancements remain identifiable through SolO and \textit{Wind}.

This evolutionary pattern agrees well with recent observational studies suggesting that microstreams contain fewer switchbacks than their near-Sun counterparts while maintaining enhanced bulk flow speeds \citep{Soni2024}. The simultaneous decrease in magnetic wave activity and persistence of the velocity structures is therefore consistent with microstreams being an evolutionary stage of SB patches, although we find through this slow Alfv\'enic stream, the energy from this dissipation is not a necessary condition to preserve these structures. 

\textbf{Future work}
It is worth noting here that not all four structures evolve identically. As discussed earlier, the left hand side of the interval provides the most reliable spacecraft-to-spacecraft mapping, with well aligned footpoints and radial energy profiles that closely follow the expected evolution. In contrast, the right most SolO structure appears to originate from a different region of the coronal hole and therefore should be interpreted more cautiously.

As discussed earlier, the present interval samples Alfv\'enic slow solar wind, and whether the same evolutionary behavior applies to fast coronal hole wind remains an open question. In a follow-up study, we will apply the same methodology to recent encounters, such as the fast wind interval studied by \citet{Bandyopadhyay2026,Badman2026}, to test the generality of the proposed evolutionary scenario and dynamical importance of the switchbacks themselves.

\section{Acknowledgement}
Parker Solar Probe was designed, built, and is now operated by the Johns Hopkins Applied Physics Laboratory as part of NASA’s Living with a Star (LWS) program (contract NNN06AA01C).
Support from the LWS management and technical team has played a critical role in the success of the {\it{Parker}} mission. 
CfA co-authors were partially supported by Parker Solar Probe project through the SAO/SWEAP subcontract 975569.
The Python Jupyter notebooks used for the data analysis are available from the corresponding author.

\appendix
\section{Additional Data Details}
\label{app:additional_Data}
The Low Frequency Receiver (LFR) provides an accurate measurement of the electron density during the times when the spacecraft is close enough to the Sun making the plasma frequency lie within the LFR frequency range and given that strong radio emissions do not overlap the same frequency range \citep{Moncuquet2020}. 
Both conditions are satisfied for the stream studied here.

\section{Additional DTW Analysis}
\label{app:additional_dtw}

Figure \ref{fig:appendix_dtw} provides an illustration of the Dynamic Time Warping (DTW) mapping applied to the \textit{Parker} and SolO normalized total energy flux longitude series. 
The connecting lines illustrate the DTW mappings, and red linkages identify regions of local compression, whereas blue linkages identify regions of local rarefaction required to align the longitude series. A vertical offset is added for visual clarity. 

\renewcommand{\thefigure}{B\arabic{figure}}
\renewcommand{\theHfigure}{B.\arabic{figure}}
\setcounter{figure}{0}

\begin{figure*}[ht!]
\centering
\includegraphics[width=0.6\textwidth]{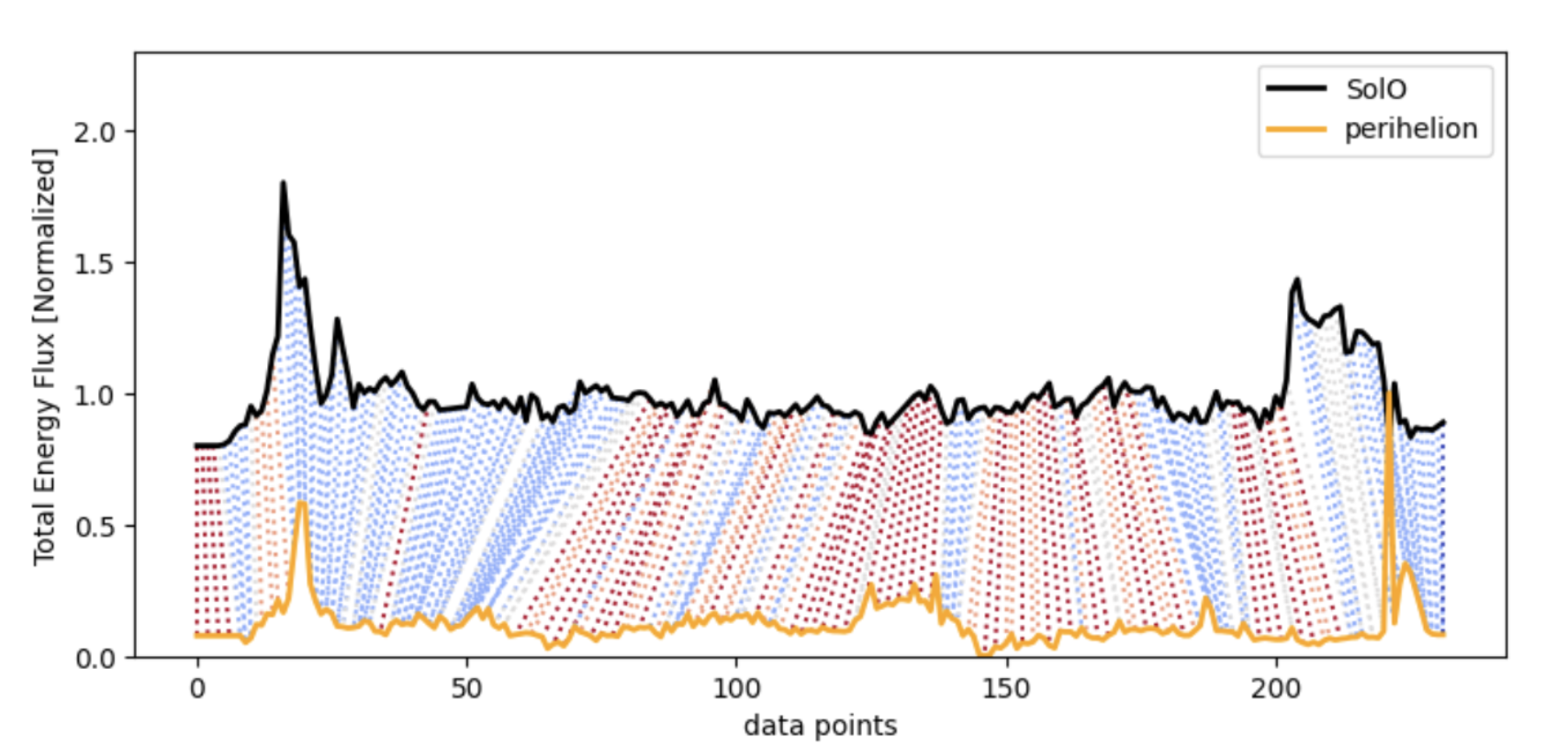}
\caption{Dynamic time warping (DTW) mapping applied to the normalized total energy flux [dimensionless] measured by {\it{Parker}} at perihelion (orange) and SolO (black). The x-axis represents the data point indices used in the DTW alignment, which are longitudinal data sampled onto a "uniform grid". Dotted connecting lines show the DTW mapping lines, with red indicating compression and blue indicating rarefaction required to align the measurements.}
\label{fig:appendix_dtw}
\end{figure*}

\renewcommand{\thefigure}{C\arabic{figure}}
\renewcommand{\theHfigure}{C.\arabic{figure}}
\setcounter{figure}{0}

\begin{figure*}[ht!]
\centering
\includegraphics[width=0.7\textwidth]{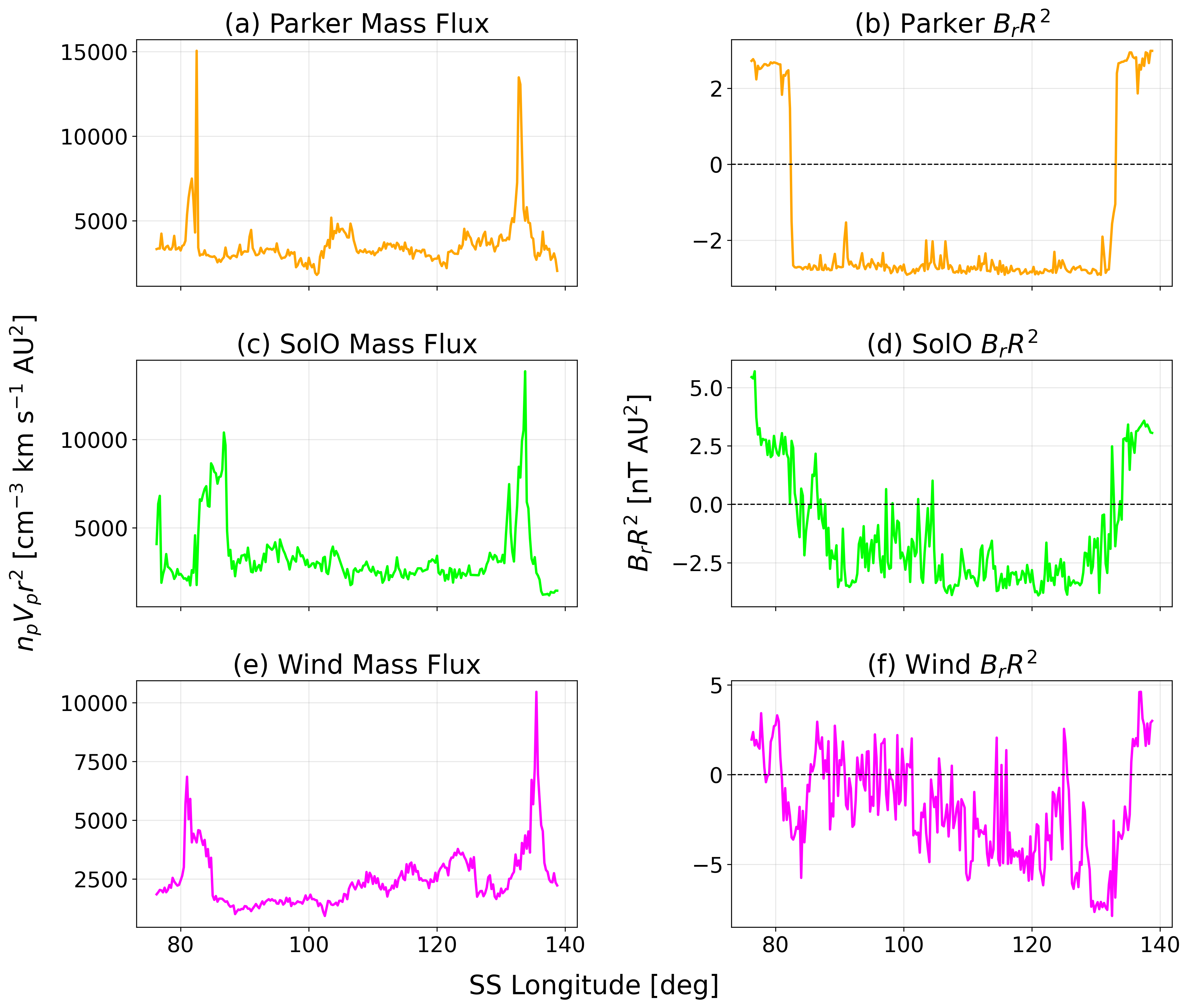}
\caption{Longitude series of the mass flux $n_p V_p r^2$ (left column), and the magnetic flux $B_R R^2$ (right column), measured by \textit{Parker} (top row), SolO (middle row), and \textit{Wind} (bottom row). These quantities are used to calculate the ratio, $g$=$n_p V_p/|B_R|$. The horizontal dashed lines in the right column indicate $B_R r^2=0$, corresponding to magnetic polarity reversals.}
\label{fig:appendix_npVpBr}
\end{figure*}

\section{Mass Flux to Magnetic Flux Ratio}
\label{app:npVpBr}

Figure \ref{fig:appendix_npVpBr} shows the mass flux $n_p V_p r^2$, and the magnetic flux $B_R r^2$, as functions of source surface longitude for \textit{Parker}, SolO, and \textit{Wind}. These quantities form the numerator and denominator, respectively, of the normalized mass flux ratio, $g$=$n_p V_p/|B_R|$, examined in the main text.

\section{Additional Table Results}
\label{app:table}

Table \ref{tab:appendix_table} summarizes the plasma and magnetic field properties of the four mapped switchback patches (orange, red, purple, and brown) observed by \textit{Parker}, SolO, and \textit{Wind}. The table includes helium abundance, plasma beta, cross helicity, magnetic fluctuation power, velocity, velocity enhancement, differential streaming, longitudinal extent, mass flux, individual energy flux components, temperature, and the temperature difference $\Delta T=T_{\max}-T_{\min}$. 

The values are reported as patch averages, with means $\pm$ one standard deviation within each patch. These standard deviations describe variability within the patches.

\renewcommand{\thetable}{D\arabic{table}}
\renewcommand{\theHtable}{D.\arabic{table}}
\setcounter{table}{0}

\startlongtable
\begin{deluxetable*}{l l l l l l }

\tablecaption{
\label{tab:appendix_table}
Patch averaged plasma and magnetic parameters for the mapped structures.
}

\tablehead{
\colhead{Parameter} &
\colhead{Spacecraft} &
\colhead{Orange} &
\colhead{Red} &
\colhead{Purple} &
\colhead{Brown}
}

\startdata
He Abundance (\%) & Parker & $3.18 \pm 0.31$ & $3.50 \pm 0.58$ & $3.13 \pm 0.29$ & $3.30 \pm 0.60$\\
 & SolO &  -- & --- & --- & ---\\
 & \textit{Wind} & $5.11 \pm 0.70$ & $4.68 \pm 0.84$ & $5.28 \pm 0.97$ & $4.27 \pm 1.08$\\
\tableline
Plasma $\beta$ & Parker & $0.10 \pm 0.01$ & $0.09 \pm 0.01$ & $0.09 \pm 0.01$ & $0.13 \pm 0.01$\\
 & SolO & $0.58 \pm 0.17$ & $0.50 \pm 0.08$ & $0.49 \pm 0.05$ & $0.59 \pm 0.09$\\
 & \textit{Wind} & $0.23 \pm 0.05$ & $0.66 \pm 0.14$ & $0.63 \pm 0.15$ & $0.75 \pm 0.27$\\
\tableline
Cross Helicity & Parker & $0.91 \pm 0.09$ & $0.86 \pm 0.15$ & $0.80 \pm 0.19$ & $0.72 \pm 0.30$\\
 & SolO & $0.68 \pm 0.32$ & $0.68 \pm 0.31$ & $0.72 \pm 0.31$ & $0.61 \pm 0.33$\\
 & \textit{Wind} & $0.53 \pm 0.33$ & $0.66 \pm 0.30$ & $0.81 \pm 0.17$ & $0.27 \pm 0.52$\\
\tableline
\shortstack[l]{Magnetic Fluctuation Power\\($\delta B^2$)} & Parker & $140508.58$ 
& $130836.09$ 
& $113530.90$ 
& $60965.20$ 
\\
 & SolO & $167.40$ 
 & $214.89$ 
 & $281.28$ 
 & $452.80$ 
 \\
 & \textit{Wind} & $3.96$ 
 & $6.69$ 
 & $6.13$ 
 & $4.50$ 
 \\
\tableline
Velocity (km s$^{-1}$) & Parker & $252.87 \pm 15.55$ & $290.69 \pm 14.00$ & $300.68 \pm 13.45$ & $231.59 \pm 8.28$\\
 & SolO & $353.01 \pm 8.77$ & $394.07 \pm 13.34$ & $499.39 \pm 11.86$ & $529.25 \pm 16.65$\\
 & \textit{Wind} & $393.30 \pm 7.07$ & $427.95 \pm 7.73$ & $418.57 \pm 6.33$ & $453.68 \pm 5.94$\\
\tableline
$\Delta V$ (km s$^{-1}$) & Parker & $54.86 \pm 41.03$ & $98.15 \pm 46.95$ & $86.15 \pm 49.80$ & $64.78 \pm 27.22$\\
 & SolO & $80.71 \pm 5.72$ & $56.12 \pm 20.36$ & $198.30 \pm 15.35$ & $100.16 \pm 33.41$\\
 & \textit{Wind} & $75.20 \pm 7.90$ & $50.80 \pm 10.11$ & $77.00 \pm 14.46$ & $69.05 \pm 15.20$\\
\tableline
\shortstack[l]{Differential \\Streaming (km s$^{-1}$)} & Parker & $102.93 \pm 24.86$ & $105.40 \pm 24.11$ & $113.93 \pm 23.97$ & $98.23 \pm 13.96$\\
 & SolO & --- & --- & --- & ---\\
 & \textit{Wind} & $3.82 \pm 2.34$ & $9.09 \pm 4.08$ & $7.51 \pm 2.71$ & $15.05 \pm 4.12$\\
\tableline
\shortstack[l]{Longitudinal \\Extent (\textdegree)} & Parker & 11.0 & 10.0 & 11.0 & 8.0\\
 & SolO & 8.5 & 11.8 & 9.25 & 5.75\\
 & \textit{Wind} & 16.2 & 10.0 & 6.25 & 4.00\\
\tableline
Mass Flux ($\mathrm{cm^{-3}\,km\,s^{-1}{AU}^2}$) & Parker & $3174.88 \pm 215.38$ & $3304.76 \pm 221.33$ & $3203.44 \pm 172.25$ & $3563.03 \pm 232.69$\\
 & SolO & $3715.34 \pm 364.71$ & $3132.15 \pm 241.78$ & $2622.90 \pm 190.10$ & $2699.64 \pm 233.15$\\
 & \textit{Wind} & $1500.88 \pm 95.27$ & $1936.14 \pm 129.56$ & $2829.44 \pm 166.38$ & $2234.42 \pm 144.61$\\
\tableline
$W_K$ (W m$^{-2}$) & Parker & $9.44 \pm 3.38$ & $13.72 \pm 6.86$ & $13.40 \pm 4.45$ & $8.80 \pm 2.20$\\
 & SolO & $17.90 \pm 6.45$ & $19.04 \pm 4.60$ & $25.82 \pm 7.51$ & $29.55 \pm 7.33$\\
 & \textit{Wind} & $9.05 \pm 2.02$ & $13.68 \pm 3.33$ & $19.50 \pm 5.30$ & $17.87 \pm 4.44$\\
\tableline
$W_G$ (W m$^{-2}$) & Parker & $48.78 \pm 5.36$ & $51.59 \pm 14.73$ & $49.24 \pm 5.76$ & $55.63 \pm 9.17$\\
 & SolO & $54.11 \pm 21.45$ & $45.59 \pm 7.66$ & $38.15 \pm 5.26$ & $39.24 \pm 7.05$\\
 & \textit{Wind} & $22.04 \pm 3.72$ & $28.43 \pm 7.06$ & $41.55 \pm 7.50$ & $32.81 \pm 7.38$\\
\tableline
$W_H$ (W m$^{-2}$) & Parker & $7.12 \pm 1.13$ & $7.75 \pm 2.49$ & $7.59 \pm 1.48$ & $7.69 \pm 1.51$\\
 & SolO & $1.26 \pm 0.42$ & $1.65 \pm 0.39$ & $2.69 \pm 0.90$ & $3.28 \pm 1.06$\\
 & \textit{Wind} & $0.31 \pm 0.13$ & $0.70 \pm 0.34$ & $0.87 \pm 0.31$ & $0.90 \pm 0.40$\\
\tableline
$W_w$ (W m$^{-2}$) & Parker & $0.93 \pm 0.16$ & $1.87 \pm 0.45$ & $1.32 \pm 0.47$ & $0.36 \pm 0.18$\\
 & SolO & $0.12 \pm 0.18$ & $0.18 \pm 0.22$ & $0.15 \pm 0.25$ & $0.98 \pm 1.13$\\
 & \textit{Wind} & $0.32 \pm 0.33$ & $0.11 \pm 0.12$ & $0.42 \pm 0.39$ & $0.49 \pm 0.45$\\
\tableline
$W_{\rm T}$ (W m$^{-2}$) & Parker & $66.27 \pm 9.57$ & $74.93 \pm 23.94$ & $71.55 \pm 11.50$ & $72.47 \pm 11.96$\\
 & SolO & $73.40 \pm 27.56$ & $66.47 \pm 12.26$ & $66.82 \pm 12.43$ & $73.06 \pm 14.81$\\
 & \textit{Wind} & $31.72 \pm 5.67$ & $43.07 \pm 10.57$ & $62.61 \pm 13.08$ & $52.16 \pm 11.94$\\
\tableline
\shortstack[l]{Temperature (MK)} & Parker & $0.65 \pm 0.063$ & $0.66 \pm 0.081$ & $0.72 \pm 0.076$ & $0.61 \pm 0.049$\\
 & SolO & $0.12 \pm 0.010$ & $0.18 \pm 0.015$ & $0.34 \pm 0.024$ & $0.40 \pm 0.032$\\
 & \textit{Wind} & $0.07 \pm 0.011$ & $0.11 \pm 0.014$ & $0.10 \pm 0.009$ & $0.13 \pm 0.014$\\
\tableline
$\Delta T$ (MK) & Parker & 0.31 & 0.39 & 0.54 & 0.46\\
 & SolO & 0.12 & 0.08 & 0.38 & 0.33\\
 & \textit{Wind} & 0.07 & 0.08 & 0.07 & 0.10\\
\enddata
\end{deluxetable*}

\bibliography{main}

\end{document}